\documentclass[apj]{emulateapj}
\usepackage{apjfonts}
\usepackage{times}

\usepackage{amsmath}
\usepackage{graphicx}
\usepackage{natbib}
\usepackage{color}

\slugcomment{Draft version}
\shorttitle{}
\shortauthors{Woo et al.}

\newcommand{\Hb}{\rm H{$\beta$}}
\newcommand{\Ha}{\rm H{$\alpha$}}

\newcommand{\mbh}{$M_{\rm BH}$}     
\newcommand{\ergs}{erg~s$^{\rm -1}$}
\newcommand{\msun}{$M_{\odot}$}       
\newcommand{\msigma}{$M_{\rm BH}-\sigma_{*}$}
\newcommand{\kms}{km~s$^{\rm -1}$}
\newcommand{\sigmastar}{$\sigma_{*}$}
\newcommand{\lam}{$\lambda$}

\newcommand{\NI}{[N {\small I}]}
\newcommand{\NII}{[N {\small II}]}
\newcommand{\SII}{[S {\small II}]}
\newcommand{\OIII}{[O {\small III}]}
\newcommand{\FeII}{Fe {\small II}}
\newcommand{\FeVII}{[Fe {\small VII}]}
\newcommand{\CaV}{[Ca {\small V}]}
\newcommand{\CaII}{Ca {\small II}}

\newcommand{\Mgb}{Mg $b$}

\newcommand{\sigmaha}{$\sigma_{\rm H\alpha}$}
\newcommand{\sigmahb}{$\sigma_{\rm H\beta}$}

\newcommand{\sigmgb}{$\sigma_{\rm Mgb}$}
\newcommand{\signomgb}{$\sigma_{\rm no Mg b}$}
\newcommand{\sigcat}{$\sigma_{\rm CaT}$}

\newcommand{\fwha}{\ensuremath{\mathrm{FWHM}_\mathrm{H{\alpha}}}}
\newcommand{\fwhb}{\ensuremath{\mathrm{FWHM}_\mathrm{H{\beta}}}}
\newcommand{\lha}{\ensuremath{L_{\mathrm{H{\alpha}}}}}
\newcommand{\lf}{\ensuremath{L_{5100}}}
\newcommand{\ledd}{$L/L_{\rm Edd}$}

\begin{document}
\title{The Black Hole Mass-Stellar Velocity Dispersion relation of Narrow-Line Seyfert 1 Galaxies}
\author{Jong-Hak Woo$^{1}$}
\author{Yosep Yoon$^{1}$}
\author{Songyoun Park$^{1}$}
\author{Daeseong Park$^{2}$}
\author{Sang Chul Kim$^{3}$}

\affil{
$^1$Department of Physics and Astronomy, Seoul National University, 
Seoul, 151-742, Republic of Korea\\
$^2$ Department of Physics and Astronomy, University of California, Irvine, CA 92697-4575, USA\\
$^3$Korea Astronomy and Space Science Institute, Daejeon, 305-348 and 
Korea University of Science and Technology (UST), Daejeon, 305-350, Republic of Korea}

\begin{abstract}
Narrow-line Seyfert 1 galaxies (NLS1s) are arguably one of the key AGN subclasses in 
investigating the origin of the black hole mass - stellar velocity dispersion (\msigma) relation 
because of their high accretion rate and significantly low \mbh. Currently, it is 
under discussion whether present-day NLS1s offset from the \msigma\ relation. 
Using the directly measured stellar velocity dispersion of 93 NLS1s at z$<$0.1,
and \mbh\ estimates based on the updated mass estimators,
we investigate the \msigma\ relation of NLS1s in comparison with broad-line AGNs. 
We find no strong evidence that the NLS1s deviates from the \msigma\ relation,
which is defined by reverberation-mapped type 1 AGNs and quiescent galaxies.
However, there is a clear trend of the offset with the host galaxy morphology, 
i.e., more inclined galaxies toward the line-of-sight have higher stellar velocity dispersion, 
suggesting that the rotational broadening plays a role in measuring 
stellar velocity dispersion based on the single-aperture spectra from the 
Sloan Digital Sky Survey.
In addition, we provide the virial factor $\log f = 0.05 \pm 0.12$ (f = 1.12), for \mbh\ estimators
based on the FWHM of \Hb, by jointly fitting
the \msigma\ relation using quiescent galaxies and reverberation-mapped AGNs. 
\end{abstract}
\keywords{galaxies: active -- galaxies: nuclei -- galaxies: Seyfert}

\section{INTRODUCTION} \label{section:intro}
The scaling relation between black hole mass and host-galaxy properties, e.g., 
the black hole mass$-$stellar velocity dispersion relation (\msigma), 
suggests a coevolution of black holes and galaxies
\citep[e.g.,][]{FerrareseMerritt2000, Gebhardt2000, HaringRix2004, McConnell2013, KormendyHo2013}, 
motivating various theoretical and observational 
studies to constrain the origin of the scaling relations and their cosmic evolution \citep{Bower2006, Croton2006, Robertson2006,Treu2007, Woo2006, Woo2008, Bennert2011b, Booth2011, Harris2012, Zhang2012, Park2014, Bennert2014}. 
Along with inactive galaxies, galaxies hosting active galactic nuclei (AGN) also seem to follow the 
\msigma\ relation with a similar slope \citep[e.g.,][]{Woo2010, Park2012, Woo2013}, indicating that the present-day galaxies show a similar
scaling relation regardless of black hole activity.

In contrast, it has been debated whether present-day narrow-line Seyfert 1 galaxies (NLS1s) 
deviate from the \msigma\ relation \citep[e.g.,][]{Mathur2001, KomossaXu2007}. As a sub-class of AGNs, 
NLS1s were initially identified by the relatively small width of the broad-component of the Balmer 
lines (FWHM $<$ 2000 \kms) and a weak \OIII-to-\Hb\ ratio (\OIII/\Hb\ $<$ 3; Osterbrock \& Pogge 1985).
Since NLS1s are believed to have small black hole masses and high Eddington ratios (Boroson 2002),
NLS1s are often considered as relatively young AGNs hosting black holes in a growing phase 
although the time evolution among various types of AGNs is highly uncertain. 
Thus, it is interesting to investigate the location of NLS1s in the \msigma\ plane in the context
of black hole-galaxy coevolution. 

A number of studies have been devoted to studying the \msigma\ relation of NLS1s over the last decade,
resulting in a controversy.
On the one hand, some studies claimed that NLS1 lie below the \msigma\ relation on average 
with smaller black hole masses at fixed stellar velocity dispersions, compared to 
the broad-line AGNs and quiescent galaxies \citep[e.g.,][]{Mathur2001, GrupeMathur2004, MathurGrupe2005a, MathurGrupe2005b, Bian2006, 
Zhou2006, Watson2007}. On the other hand, other studies reported that the NLS1s are generally 
on the \msigma\ relation \citep[e.g.,][]{WangLu2001, KomossaXu2007}. 
The fundamental limitation of the aforementioned studies is the fact that stellar velocity dispersions were
not directly measured. Instead, the width of the narrow \OIII\ emission line at 5007\AA\ was used as 
a surrogate for stellar velocity dispersion, based on the empirical correlation between \OIII\ width and stellar velocity dispersion \citep{Nelson2000}, although there is a considerably large scatter between them. 
If the ionized gas in the narrow-line region follows the gravitational potential of the host-galaxy,
then \OIII\ line width can be substituted for stellar velocity dispersion. However for individual objects
the uncertainty of this substitution is very large as shown by the direct comparison between \OIII\ width 
and the measured stellar velocity dispersion \citep[e.g.,][]{Woo2006, Xiao2011}.
Moreover, the \OIII\ line often suffers from the effect of outflow, manifesting an 
asymmetric line profile and a strong blue-shifted wing component \citep[e.g.][]{Boroson2005, BaeWoo2014}. 
In this case, the width of the \OIII\ line will become much broader than 
stellar velocity dispersion, if the blue wing is not properly corrected for. 
In fact, \cite{KomossaXu2007} showed that when the blue wing component is removed 
in measuring the width of the \OIII\ line, the inferred stellar velocity dispersion from \OIII\ 
becomes smaller, hence the NLS1 show a consistent \msigma\ relation compared to broad-line AGNs.

The solution to this decade-long debate is to investigate the locus of NLS1s in the \msigma\ plane, using {\it directly measured} stellar velocity dispersion. 	
Although, measuring stellar velocity dispersion of AGN host galaxies is difficult due to the presence of
strong AGN features, i.e., power-law continuum, \FeII\ emission, and broad emission lines,
it is possible to measure stellar velocity dispersion if high quality spectra are available 
as demonstrated in a number of studies \citep[e.g.,][]{Woo2006,GH2006,Woo2010,Hiner2012, 
Woo2013}. 
In this paper, we present the direct stellar velocity dispersion measurements and estimates 
of black hole masses for a sample of 93 NLS1s at z $< 0.1$ 
selected from Sloan Digital Sky Survey Data Release 7 (SDSS DR7) \citep{Abazajian2009},
in order to investigate the \msigma\ relation of NLS1s. 
We describe the sample selection and properties in Section 2, and the analysis including mass determination and stellar velocity dispersion measurements in Section 3.  Section 4 presents the results, followed by discussion in Section 5, and summary and conclusions in Section 6. 
Throughout the paper, we adopt a cosmology of $H_{\rm 0}= 70$ km  s$^{-1}$ Mpc$^{-1}$, 
$\Omega_{\Lambda}=0.7$ and $\Omega_{\rm m}=0.3$.

\section{Sample and Data}

\subsection{Sample selection}

NLS1s are generally defined with two criteria: (1) the full-width-at-half-maximum (FWHM) of broad 
component of the Balmer lines $<$ 2000 \kms, and (2) the line flux ratio 
\OIII/\Hb\ $<$ 3 \citep{Osterbrock1985, Goodrich1989}. Additional characteristics of NLS1s 
include strong \FeII\ emission \citep{Osterbrock1985}, 
high Eddington ratio and soft X$-$ray emission \citep{Leighly1999, Grupe2004, McHardy2006}. 
In this study, we selected a sample of NLS1s from SDSS DR7 \citep{Abazajian2009},
based on the width of Balmer lines and the \OIII/\Hb\ flux ratios.
First, we selected NLS1 candidates by limiting the width of H$\beta$ to 500$-$2500 km  s$^{-1}$, 
using the SpecLine class in the SDSS Query tool (http://casjobs.sdss.org).
Since the line width measurements from the SDSS pipeline is not precise,
we used a wider width range than the conventional definition 
for the initial selection, obtaining 4,252 NLS1 candidates at z $<$ 0.1.

Second, using this initial sample, we performed a multi-component spectral decomposition 
analysis for each galaxy, to properly measure the width of the broad component of the 
Balmer lines.
In the fitting process, we included multiple components, namely, 
featureless AGN continuum, stellar population model, 
and \FeII\ emission component, using an IDL-based spectral decomposition code \citep[see][]{Woo2006, Park2012, Park2014}.
By subtracting the linear combination of featureless AGN continuum, stellar component and \FeII\ emission, we obtained emission line spectra and fit the broad and narrow emission lines 
(see section 3.1 for the detailed fitting process).
Based on the measurements from the fitting process, we finalized a sample of 464 NLS1s,
that satisfy the aforementioned two criteria by limiting the FWHM of broad \Ha\ 
between 800 and 2200 \kms\ and the line flux ration \OIII/\Hb\ less than 3.

Among these objects, we measured and collected the stellar velocity dispersion for 93 NLS1s.
For 63 objects, we were able to directly measure stellar velocity dispersion using the SDSS spectra
(see Section 3.5) while for 30 objects we obtained the stellar velocity dispersion measurements 
from \cite{Xiao2011}. Thus, using this sample of 93 NLS1s, we investigate the properties of NLS1 and the \msigma\ relation.
Note that the distribution of NLS1 properties (i.e., \OIII/\Hb\ ratio, \FeII/\Hb\ ratio, \Ha\ luminosity and width)
of the final sample of 93 objects is similar to that of the initial sample of 464 objects,
suggesting that we may treat the final sample as a random subsample of NLS1 galaxy population.

\subsection{Sample properties}

\begin{figure}
\includegraphics[width = 0.48\textwidth]{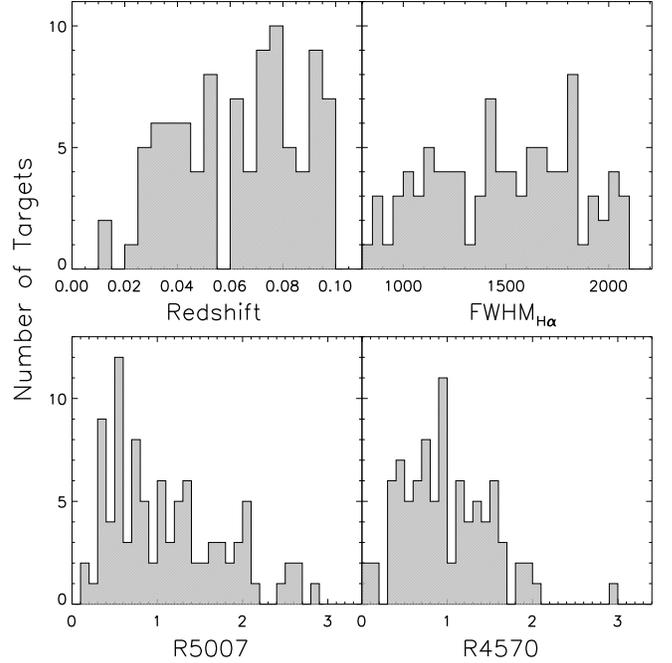}
\caption{Histogram of the NLS1s showing the distribution of redshift, \fwha, R5007 and R4570.
The \Hb\ total flux is used to obtain R5007 and R4570. 
\label{fig:histogram}}
\end{figure} 

Figure 1 presents the distributions of redshift and the width of \Ha\ of the final sample
(top panels). 
To demonstrate the weak \OIII\ emission and strong \FeII\ emission of the sample 
as the characteristic features of NLS1s \citep{Osterbrock1985, Goodrich1989, Veron2001}, 
we also present the distribution of the flux ratio \OIII/\Hb\ (R5007) and \FeII/\Hb\ (R4570) ratios in Figure 1 (bottom panels). 
Since the \Hb\ is relatively weak and the decomposition of the broad and narrow components of \Hb\ 
is uncertain, we used the total \Hb\ flux to compare with \OIII\ and \FeII\ fluxes. 

In the case of the \OIII\ strength (R5007), all galaxies in our sample show low 
\OIII/\Hb\ ratio ($<$3), with a median 1.05 and a mean 1.14. 
The \FeII\ strength (R4570), defined by the line flux ratio of \FeII\ emission 
integrated over the 4434$-$4684\AA\ region, to \Hb\ \cite[e.g.,][]{Veron2001}, is also high 
with a mean 1.06, as similarly found by other NLS1 studies
(for dependence on the R4570 index, see \S 4.1).
For example, \cite{Zhou2006} used the broad component of \Hb\ to compare
with \FeII\ and reported the mean R4570 as $\sim$0.82, while \cite{Xu2012} adopted 
the total flux of  \Hb\ and found the mean R4570 $\sim$0.7. 

The \mbh\ estimated with the line dispersion of broad component of \Ha\ ranges over an order of magnitude,
i.e., log \mbh/\msun = 5.84$-$7.38 with a mean 6.72, 
which is comparable to that of the previous NLS1 \msigma\ relation studies
\citep{Grupe2004, KomossaXu2007}. 
The Eddington ratio of our NLS1s ranges from 5\% to the Eddington limit
with a mean of 0.2-0.3, depending on the mass estimates.

\section{Analysis}

\subsection{Multi-component fitting of the emission lines}

We performed multi-components spectral fitting analysis in two separate spectral ranges:
 \Hb\ region (4400$-$5580\AA) and \Ha\ region (6500$-$6800\AA).  
For the \Hb\ region, we followed the procedure given by our previous studies 
\citep[][see also Barth et al. 2013]{Woo2006, McGill2008, Park2012, Park2014, Woo2014}. 
After converting all spectra to the rest frame, we modeled the observed spectra with three components, 
i.e., featureless AGN continuum, host-galaxy starlight, and \FeII\ emission blends, by respectively using a single power law continuum, a stellar population model
based on the SED templates from \cite{BC03}, 
and an \FeII\ template from \citep{BG92}. 
The best continuum model was determined in the regions 4430$-$4600\AA\ and 5080$-$5550\AA, where \FeII\ emission dominates. We simultaneously fitted all 3 components, 
using the nonlinear Levenberg-Marquardt least-squares fitting routine $mpfit$ \citep{Markwardt2009} in IDL.
After subtracting the featureless AGN continuum and host-galaxy starlight 
from the raw spectra, emission line fitting for \Hb, \OIII\ \lam4959 and \OIII\ \lam5007 
 was carried out for this region. 
Since \OIII\ \lam5007 shows often complex profile such as velocity shift of \OIII\ core and asymmetry \citep{KomossaXu2007},
we decomposed the \OIII\ line into a narrow core and a broad base. 
If \OIII\ has the broad base which tends to show blue-asymmetric (blue wing), 
the \OIII\ is fitted with double Gaussian components.
On the other hand, if the \OIII\ profile is symmetric or of the S/N is low, 
the \OIII\ is fitted with a single Gaussian component. 
Then, the best-fit model of the \OIII\ \lam5007 line was used to model \OIII\ \lam4959 and \Hb\ narrow component by assuming that these narrow lines have the same widths.
The flux ratio of the \OIII\ \lam4959 to the \OIII\ \lam5007 was assumed to be 1:3, 
while the height of the \Hb\ narrow component was set as free parameter. 
Next, we fitted the \Hb\ broad component with a single Gaussian component since
the S/N of \Hb\ is typically lower than \OIII.

\begin{figure}
\includegraphics[width = 0.49\textwidth]{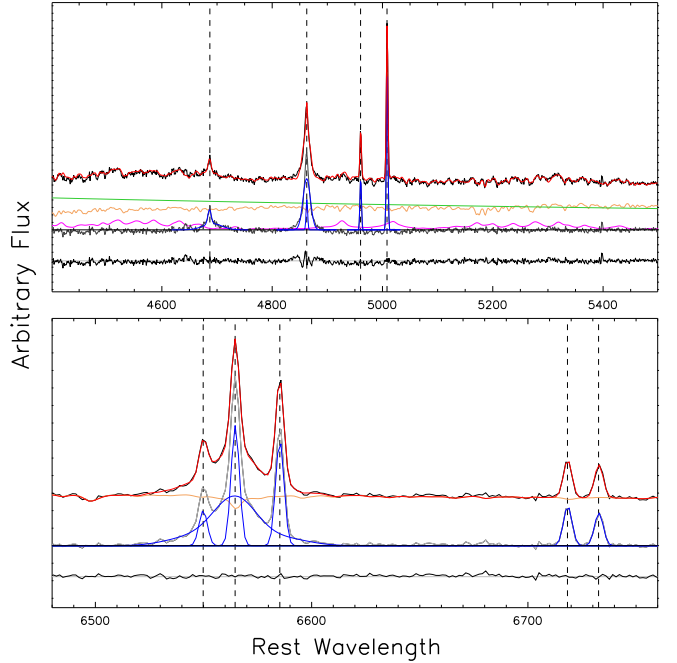}
\caption{Example of multi-component spectral fitting for the \Hb\ (top) and \Ha\ region (bottom). 
In both panels, raw spectra (black) and the best-fit of total models (red) are illustrated in upper part,
and the best-fit of emission lines (blue), power-law continuum (green), host-galaxy spectrum (orange) 
are shown in the bottom part. 
For the \Hb\ region, Fe II blend model (magenta) is presented additionally.
The residuals (black) are arbitrarily shifted downward to clarify.
\label{fig:fitting}}
\end{figure}

For the \Ha\ region, we did not subtract \FeII\ emission because \FeII\ is relatively 
weak in this spectral range. 
First, we fit the host-galaxy continuum using two spectral regions 6400$-$6460\AA\ and 6740$-$6800\AA\ for determining the best model, where no other emissions are present. 
After subtracting the stellar features,
we fitted \SII\ \lam6716 and \SII\ \lam6731, respectively with a single Gaussian component.
We assumed that the widths of \SII, \NII, and the \Ha\ narrow component are
the same, and used the width of the \SII\ for fitting \NII\ and the narrow \Ha,
if the spectral quality is high (S/N of \SII\ $> 25$). 
For low S/N targets, the width of \SII\ is not reliable and we 
fitted the \Ha\ narrow component and the \NII\ doublet with a single Gaussian 
model, without using the best-fit of the \SII\ line.  
The flux ratio between \NII\ \lam6548 and \NII\ \lam6583 is assumed as 1/3. 
For the \Ha\ broad component, Gauss-Hermitian series were used to model 
the \Ha\ profile as done by \cite{McGill2008}. 
Figure~\ref{fig:fitting} presents an example of the multicomponent fitting.

We estimated the uncertainty of the \Ha\ luminosity based on the S/N of the line flux. 
In the case of the line widths, we performed Monte Carlo simulations by randomizing the flux per pixel
using the flux noise. For a set of 100 simulated spectra, we repeated spectral decomposition, measured 
the line width, and adopted the 1-sigma dispersion of the distribution as the uncertainty of the line widths for each object. The estimated uncertainties are included in Table 1. 

\subsection{Black hole mass}

Black hole mass can be determined based on the virial theorem:
\begin{equation}\label{eq:rm_mass}
M_{\rm BH}=f\frac{V^2R_{\rm BLR}}{G}\ 
\end{equation}
where $V$ is the velocity of the broad-line region (BLR) gas, $R_{\rm BLR}$ is the BLR size, and G
is the gravitational constant \citep{Peterson2004}. 
Generally, either the second moment (line dispersion; \sigmahb) or the FWHM of the \Hb\ line (\fwhb)        
is used for the velocity of the BLR gas. Along with each velocity measurements, 
a virial factor f is needed for mass determination. 
The determination of the average virial factor, respectively, for \sigmahb\ and \fwhb\
can be found in Appendix, where we derived the virial factor by comparing 
the reverberation-mapped AGNs and quiescent
galaxies in the \msigma\ plane.

Instead of directly measuring the size of BLR by reverberation mapping, 
which requires a long-term spectroscopic monitoring, 
an empirical size-luminosity relation \cite[e.g.,][]{Kaspi2000, Kaspi2005, Bentz2009a, Bentz2013} has been used for \mbh\ estimates. 
We used the size$-$luminosity relation from \cite{Bentz2013}, and derive 
the \mbh\ estimator as follows, 
\begin{equation}
M_{\rm BH} = f \times 10^{6.819} 
\left(\frac{\rm \sigma_{H\beta}}{10^3~\rm {km~s^{-1}}}\right)^{2}
\left(\frac{\lambda L_{5100}}{10^{44}~{\rm erg~s^{-1}}}\right)^{0.533}~{\rm M}_{\odot}~.
\end{equation}

For our NLS1s, the width of the \Ha\ line is better determined that that of 
the \Hb\ lines since \Hb\ often have much lower S/N. Thus, we used the measurement of \Ha\ line width and luminosity for \mbh\ estimation, using the following two relations \citep{GH2005}:
\begin{equation}
\fwhb = (1.07 \pm 0.07) \times 10^3 \left(\fwha \over 10^3~\rm{km~s^{-1}}\right)
^{(1.03 \pm 0.03)}~{\rm{km~s^{-1}}}
\end{equation}
\begin{equation}
\lha = (5.25 \pm 0.02) \times 10^{42} 
\left( \frac{\lambda\lf}{10^{44}~{\rm erg~s^{-1}}}\right)
^{(1.157 \pm 0.005)}~{\rm erg~s^{-1}} 
\end{equation}
Assuming the \Hb\ and \Ha\ have the same line profile (i.e., FWHM = 2 $\sigma$), 
we also converted \sigmaha\ to \sigmahb\ using Eq. 3.
To test the validity of Eq. 3 for our NLS1, we compared the line width
of \Hb\ and \Ha\ using a subsample of 41 NLS1s, for which the S/N ratio of \Hb\
is larger than 20 so that we could obtain reliable emission line fitting results. 
We find that the relation between \Ha\ and \Hb\ of NLS1s is consistent with that of 
reported by \citet{GH2005}, with a slight offset $0.041 \pm 0.009$ from the equation (3).
For comparing \lf\ with \Ha\ luminosity, we used all NLS1s in our sample,
for which \lf\ was measured from a power-law component in the multi-component fitting 
process. As shown in Figure 3, the relation between \lf\ and \Ha\ luminosity is close 
to Equation (4), with a slight offset $0.077 \pm 0.202$.
This result suggests that using the conversion equation is acceptable for NLS1s
and that the multi-component fitting results are reasonable, 
although a proper comparison is difficult due to the limited dynamical range of
the NLS1 sample compared to that of \cite{GH2005}. 
Note that we used a Gauss-Hermite series for the broad \Ha\ component, and a single Gaussian 
model for the broad \Hb\ component (due to low S/N ratio), while \cite{GH2005} used a multicomponent Gaussian models for both \Ha\ and \Hb. The difference of the fitting model may be partly 
responsible for the slight systematic offset.

\begin{figure}
\includegraphics[width = 0.48\textwidth]{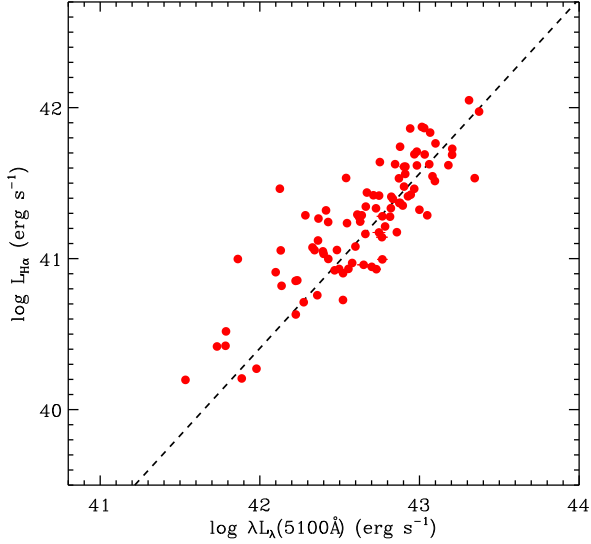}
\caption{Comparison between \lf\ and \Ha\ luminosity of the NLS1s. The dashed line 
represents the derived relation by cite{GH2005}. 
\label{fig:HaHb}}
\end{figure}

We derived a black hole mass estimators by combining aforementioned scaling relations
as:.
\begin{equation}
M_{\rm BH} = f \times 10^{6.544} \left(\frac{\lha}{10^{42}~{\rm erg~s^{-1}}}\right)^{0.46}
\left(\frac{\rm FWHM_{H\alpha}}{10^3~\rm {km~s^{-1}}}\right)^{2.06}~{\rm M}_{\odot}~ ,
\end{equation}

\begin{equation}
M_{\rm BH} = f \times 10^{6.561} \left(\frac{\lha}{10^{42}~{\rm erg~s^{-1}}}\right)^{0.46}
\left(\frac{\rm \sigma_{H\alpha}}{10^3~\rm {km~s^{-1}}}\right)^{2.06}~{\rm M}_{\odot}~ .
\end{equation}
We adopted log f = 0.05$\pm$0.12 (f = 1.12) for FWHM-based \mbh\, 
while we used log f = 0.65$\pm$0.12 (f = 4.47) for $\sigma$-based \mbh\ (see Appendix for
detailed discussion).

\subsection{Stellar velocity dispersion}

Directly measuring stellar velocity dispersions is a key 
to determine the location of NLS1 on the \msigma\ plane. 
To investigate the systematic uncertainties of the stellar velocity dispersion measurements, 
we measured \sigmastar\ in three spectral regions:
(1) \Mgb-Fe region (5000-5430\AA), which includes strong absorption lines, i.e., 
\Mgb\ triplet (5069, 5154, 5160\AA) and Fe (5270, 5335\AA) lines (hereafter \sigmgb); 
(2) \Mgb-Fe region (5000-5430\AA) excluding the \Mgb\ triplet (hereafter \signomgb);
and (3) \CaII\ region (8400-8800\AA), where the \CaII\ triplet (8498, 8542, 8662\AA) 
is a strong feature (hereafter \sigcat). 
The line strength of the \Mgb\ triplet is much higher in the composite spectra 
of massive elliptical galaxies than in the nearby stars, hence, the template mismatch due to 
the $\alpha-$element enhancement can potentially cause a systematic bias 
in measuring \sigmastar, although this effect is not significant for
late-type host galaxies \citep{Barth2002, Barth2003, Woo2004, Woo2005}. 
In the case of the \CaII\ triplet region, AGN contamination (e.g., \FeII\ emission) is relatively weaker than the \Mgb\ region, while the residual of sky emission lines is often present and the quality of spectra is generally lower than that of the \Mgb\ region.
Thus, as a consistency check, we measured stellar velocity dispersion using three
different spectral regions (see similar investigation by Greene et al. 2005).
We find that three measurements are consistent, showing that the effect of the Mg abundance is negligible (see below). 

We corrected for the SDSS spectral resolution by subtracting the instrumental resolution
from the measured stellar velocity dispersion in quadrature. Instead of using a mean constant resolution $\sim$ 70 \kms, 
which is often adopted in the literature, we calculated the mean instrumental resolution 
in the corresponding fitting ranges for each object, using the spectral resolution fits file provided
by SDSS DR7. For example, we used the spectral range 
5000-5430 \AA\ to calculate the mean instrumental resolution for the \Mgb-Fe region,
which is $\sim$55-56 \kms. Compared to the instrumental resolution, the stellar lines of the objects 
that we mesured stellar velocity dispersions are well resolved.

After masking out AGN narrow emission lines \citep[e.g., \FeVII\ $\lambda$5160, 
\NI\ $\lambda$5201, \CaV\ $\lambda$5310;][]{VandenBerk2001}, 
we measured \sigmastar\ by using both the penalized pixel-fitting (pPXF) method 
\citep{cap04} and a Python-based code based on the algorithm by \citet{Marel1994}.
We used stellar velocity templates from INDO$-$US stellar library,
which includes various spectral type giant stars
with a range of metallicity ([Fe/H] = -0.49 $-$ 0.18) \citep{Valdes2004}.
Low order polynomials were used to fit the broad curvature in the spectra   
after masking out the narrow emission lines and bad spectral regions. 
After intense tests with various polynomial orders and templates for each target,
we adopted the mean of the measurements based on each polynomial order and 
each spectral range with a different mask-out region, as a final measurement of \sigmastar.

\begin{figure}
\includegraphics[width = 0.48\textwidth]{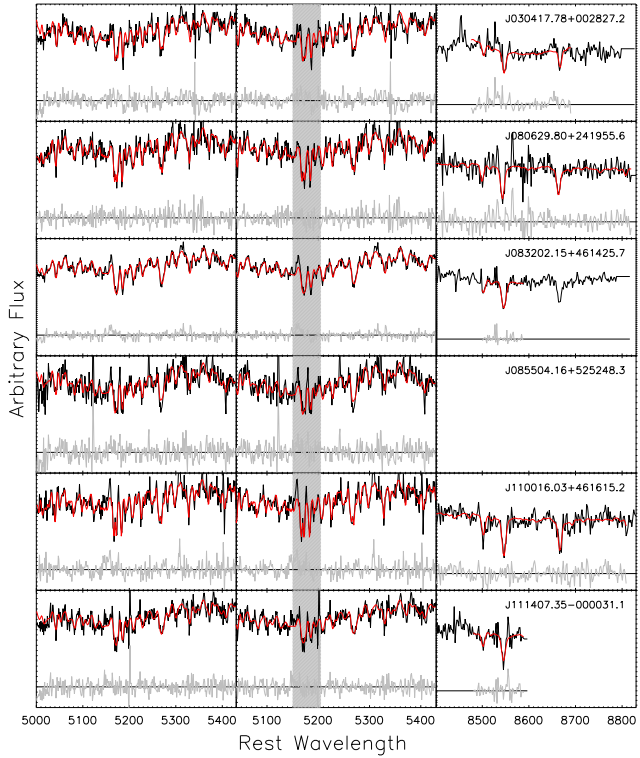}
\caption{Examples of stellar velocity dispersion fitting using the Mg b-Fe region (left)
by excluding the Mgb triplet line (middle), and the Ca II region (right). 
The observed spectrum (black line) is overplotted with the best-fit model (red line)
in each panel while the residual of the fit (gray) is shown at the bottom.
}
\end{figure} 

\begin{figure}
\includegraphics[width = 0.48\textwidth]{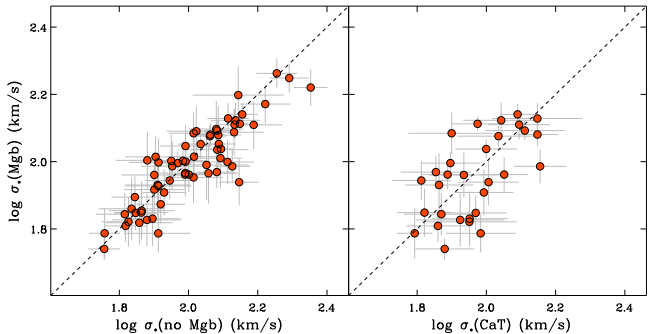}
\caption{Comparisons between the measurements of stellar velocity dispersion. 
The relations of \sigmgb\ with \signomgb\ for 93 NLS1s (left) and 
with \sigcat\ for 34 NLS1s (right) are illustrated.
\label{fig:SVD}}
\end{figure} 

In this process, we measured \sigmgb\ for 62 NLS1s, that show strong enough stellar lines.
Among them, we were able to measure \sigcat\ for 34 NLS1s, while 
we could not measure the \sigmastar\ from the \CaII\ triplet for the other objects, 
since the SDSS spectral range does not cover the rest-frame \CaII\ triplet region for targets at  z $> 0.082$,
or the strength of the \CaII\ triplet is too weak to measure \sigmastar\ (see Figure 4). 
As a consistency check, we compared the \sigmgb\ with \signomgb\ and \sigcat\
in Figure~\ref{fig:SVD}. The \sigmgb\ is slightly higher by a few percent 
(0.015 dex) than the \signomgb, and the rms scatter is 0.06 dex. 
This result confirms the \sigmgb\ is 
consistent with the \signomgb\ and indicates the influence of the \Mgb\ triplet 
is marginal in measuring the \sigmastar\ of the host galaxies of NLS1s.
The comparison between \sigcat\ and \sigmgb\ shows slightly larger scatter (0.10 dex),
but the average offset is still close to zero (i.e., 0.014 dex), suggesting that
the \sigmgb\ is consistent with the \sigcat. Based on these results without strong bias 
among the measurements from various spectral regions, 
we adopted \sigmgb\ as the final measurements.
As a consistency check, we compared our measurements with SDSS DR7 values.
We found stellar velocity dispersion measurements for 5 objects from SDSS DR7,
which are consistent with our measurements within the measurement uncertainties.

Among 93 NLS1s, 30 NLS1s were studied previously by \cite{Xiao2011}, 
who measured \sigmastar\ based on high quality spectra with higher spectral resolution 
obtained with the Keck Echellette Spectrograph and Imager (ESI) and the Magellan Echellette (MagE).
Thus, including the \sigmastar\ measurements of 30 NLS1s from \cite{Xiao2011},
we have a total of 93 measurements of the \sigmastar.
We note that the 3\arcsec\ SDSS fiber size is larger than the slit size adopted
by \cite{Xiao2011}. Thus, the SDSS spectra represent a larger physical scale of 
the host galaxies than the Keck spectra of \cite{Xiao2011},
and may show larger influence of rotational broadening. 
However, it is difficult to perform a direct comparison between SDSS-based and Keck-based
measurements due to the fact that most of 30 NLS1s studied by \cite{Xiao2011}
have smaller velocity dispersion than the SDSS instrumental resolution. 
We found only one object among 30 NLS1s, for which both SDSS-based and Keck-based stellar velocity dispersion
measurements are available and show consistency ($75\pm10$ vs. $71\pm5$ \kms).  

\subsection{Morphology classification}

For galaxies with a rotating stellar disk, the line-of-sight stellar velocity dispersion can be 
overestimated due to rotational broadening \citep{Bennert2011a, Harris2012, Kang2013},
therefore it is important to correct for the rotation effect in measuring \sigmastar. 
Since the rotating disk is common among late-type galaxies and the ratio between
rotation velocity and velocity dispersion is typically higher in late-type galaxies
than in early-type galaxies,
the effect of the rotational broadening is expected to be stronger for late type galaxies, 
particularly for more inclined galaxies toward the line-of-sight.

To investigate the rotation effect on the \msigma\ relation, 
we classified our NLS1s into early and late type galaxies, using the SDSS colors and 
the presence of a disk. For late-type galaxies, we further divided them
into two groups: more face-on and more edge-on galaxies based on the inclination of 
the disk. The inclination angle is determined from the minor-to-major axial ratio of the disk 
as $i=sin^{-1}q$, where $i$ is the inclination angle of the galactic disk to the line of sight 
(i.e., $i=0^{\circ}$ for an edge-on disk) and $q$ is the ratio of the minor to major axes of the disk. 
We classified our sample with $q > 0.5$ (i.e., $i > 30^{\circ}$) as face-on galaxies, 
and the others with $q < 0.5$ (i.e., $i < 30^{\circ}$) were classified as edge-on galaxies.
As a result, 93 NLS1 galaxies 
were divided into 35 early type galaxies and 58 late type galaxies which were further divided into 48 face-on and 10 edge-on late types.

\LongTables
\begin{deluxetable*}{lcccc cccc}
	\tablecolumns{8}
	\tablewidth{0pc}
	\tablecaption{NLS1}
	\tablehead{
		\colhead{Name}&
		\colhead{z}&
		\colhead{$\rm log\ L_{\rm H\alpha}$}&
		\colhead{$\sigma_{\rm H\alpha}$}&
		\colhead{$\rm FWHM_{\rm H\alpha}$}&
		\colhead{$\rm log\ M_{BH}(\sigma_{\rm H\alpha})$}&
		\colhead{$\rm log\ M_{BH}(\rm FWHM_{\rm H\alpha})$}&
		\colhead{$\sigma_{*}$}&
		\colhead{S/N}
		\\
		\colhead{}&
		\colhead{}&
		\colhead{($\rm erg\ s^{-1}$)}&
		\colhead{($\rm km\ s^{-1}$)}&
		\colhead{($\rm km\ s^{-1}$)}&
		\colhead{($\rm M_{\odot}$)}&
		\colhead{($\rm M_{\odot}$)}&
		\colhead{($\rm km\ s^{-1}$)}&
		\colhead{}
		\\
		\colhead{(1)}&
		\colhead{(2)}&
		\colhead{(3)}&
		\colhead{(4)}&
		\colhead{(5)}&
		\colhead{(6)}&
		\colhead{(7)}&
		\colhead{(8)}&
		\colhead{(9)}
	}
	\startdata
J010409.16+000843.6   &  0.071  &   41.29 $\pm$ 0.01    &    701 $\pm$ 11     &   1375 $\pm$ 28     &   6.57 $\pm$ 0.02     &  6.55  $\pm$ 0.02     &     66 $\pm$ 16           &   15   \\
J030417.78+002827.2   &  0.045  &   41.40 $\pm$ 0.01    &    728 $\pm$ 11     &   1248 $\pm$ 19     &   6.65 $\pm$ 0.01     &  6.51  $\pm$ 0.01     &     88 $\pm$ 8            &   30   \\
J073106.86+392644.5   &  0.048  &   41.06 $\pm$ 0.01    &    662 $\pm$ 7      &   1185 $\pm$ 19     &   6.41 $\pm$ 0.01     &  6.31  $\pm$ 0.01     &     72 $\pm$ 14           &   19   \\
J073714.28+292634.1   &  0.080  &   41.46 $\pm$ 0.01    &    966 $\pm$ 26     &   1553 $\pm$ 36     &   6.93 $\pm$ 0.03     &  6.74  $\pm$ 0.03     &    102 $\pm$ 12           &   19   \\
J080253.18+130559.6   &  0.095  &   42.05 $\pm$ 0.01    &   1072 $\pm$ 8      &   1903 $\pm$ 21     &   7.30 $\pm$ 0.01     &  7.19  $\pm$ 0.01     &     97 $\pm$ 17           &   24   \\
J080538.22+244214.8   &  0.099  &   41.61 $\pm$ 0.02    &    784 $\pm$ 14     &   1242 $\pm$ 94     &   6.81 $\pm$ 0.02     &  6.61  $\pm$ 0.02     &    102 $\pm$ 24           &   11   \\
J080801.75+381935.3   &  0.041  &   40.86 $\pm$ 0.01    &    896 $\pm$ 21     &   1683 $\pm$ 41     &   6.59 $\pm$ 0.02     &  6.53  $\pm$ 0.02     &    100 $\pm$ 12           &   20   \\
J081718.55+520147.7   &  0.039  &   41.06 $\pm$ 0.01    &    842 $\pm$ 12     &   1486 $\pm$ 32     &   6.62 $\pm$ 0.01     &  6.51  $\pm$ 0.01     &     68 $\pm$ 14           &   18   \\
J082007.81+372839.6   &  0.082  &   41.42 $\pm$ 0.01    &   1085 $\pm$ 61     &   1661 $\pm$ 92     &   7.02 $\pm$ 0.05     &  6.78  $\pm$ 0.05     &    141 $\pm$ 17           &   22   \\
J083202.15+461425.7   &  0.046  &   41.42 $\pm$ 0.01    &   1026 $\pm$ 20     &   1646 $\pm$ 42     &   6.97 $\pm$ 0.02     &  6.77  $\pm$ 0.02     &    128 $\pm$ 5            &   38   \\
J083741.94+263344.1   &  0.076  &   41.34 $\pm$ 0.01    &   1025 $\pm$ 44     &   1767 $\pm$ 131    &   6.93 $\pm$ 0.04     &  6.80  $\pm$ 0.04     &    105 $\pm$ 16           &   17   \\
J083949.65+484701.4   &  0.039  &   41.56 $\pm$ 0.01    &    904 $\pm$ 9      &   1495 $\pm$ 14     &   6.92 $\pm$ 0.01     &  6.75  $\pm$ 0.01     &    112 $\pm$ 6            &   42   \\
J084927.36+324852.8   &  0.064  &   41.64 $\pm$ 0.01    &   1235 $\pm$ 23     &   2045 $\pm$ 33     &   7.23 $\pm$ 0.02     &  7.07  $\pm$ 0.02     &    137 $\pm$ 11           &   25   \\
J085504.16+525248.3   &  0.089  &   41.86 $\pm$ 0.01    &    889 $\pm$ 20     &   1540 $\pm$ 32     &   7.04 $\pm$ 0.02     &  6.92  $\pm$ 0.02     &    103 $\pm$ 10           &   20   \\
J092438.88+560746.8   &  0.026  &   41.00 $\pm$ 0.01    &    899 $\pm$ 18     &   1723 $\pm$ 38     &   6.66 $\pm$ 0.02     &  6.62  $\pm$ 0.02     &    146 $\pm$ 5            &   39   \\
J093638.69+132529.6   &  0.090  &   41.41 $\pm$ 0.01    &   1025 $\pm$ 44     &   1916 $\pm$ 209    &   6.96 $\pm$ 0.04     &  6.91  $\pm$ 0.04     &    102 $\pm$ 11           &   17   \\
J094153.41+163621.0   &  0.052  &   41.05 $\pm$ 0.01    &   1005 $\pm$ 17     &   2078 $\pm$ 39     &   6.78 $\pm$ 0.02     &  6.81  $\pm$ 0.02     &    101 $\pm$ 11           &   15   \\
J095848.67+025243.2   &  0.079  &   41.07 $\pm$ 0.01    &   1004 $\pm$ 30     &   1710 $\pm$ 51     &   6.79 $\pm$ 0.03     &  6.65  $\pm$ 0.03     &    117 $\pm$ 11           &   18   \\
J100854.93+373929.9   &  0.054  &   41.97 $\pm$ 0.01    &   1010 $\pm$ 20     &   1750 $\pm$ 52     &   7.21 $\pm$ 0.02     &  7.08  $\pm$ 0.02     &    105 $\pm$ 9            &   38   \\
J102532.09+102503.9   &  0.046  &   41.29 $\pm$ 0.01    &    930 $\pm$ 9      &   1615 $\pm$ 20     &   6.82 $\pm$ 0.01     &  6.70  $\pm$ 0.01     &    111 $\pm$ 8            &   26   \\
J103103.52+462616.8   &  0.093  &   41.86 $\pm$ 0.01    &   1029 $\pm$ 15     &   1806 $\pm$ 30     &   7.17 $\pm$ 0.01     &  7.06  $\pm$ 0.01     &    169 $\pm$ 16           &   21   \\
J103751.81+334850.1   &  0.051  &   40.82 $\pm$ 0.01    &   1053 $\pm$ 49     &   1832 $\pm$ 72     &   6.71 $\pm$ 0.04     &  6.59  $\pm$ 0.04     &     94 $\pm$ 12           &   18   \\
J104153.59+031500.6   &  0.093  &   41.69 $\pm$ 0.01    &   1155 $\pm$ 20     &   1940 $\pm$ 35     &   7.20 $\pm$ 0.02     &  7.04  $\pm$ 0.02     &    126 $\pm$ 21           &   18   \\
J105600.39+165626.2   &  0.085  &   41.48 $\pm$ 0.01    &    996 $\pm$ 24     &   1821 $\pm$ 42     &   6.97 $\pm$ 0.02     &  6.89  $\pm$ 0.02     &    126 $\pm$ 15           &   20   \\
J110016.03+461615.2   &  0.032  &   40.91 $\pm$ 0.01    &    835 $\pm$ 11     &   1646 $\pm$ 20     &   6.55 $\pm$ 0.01     &  6.54  $\pm$ 0.01     &     68 $\pm$ 6            &   24   \\
J111253.12+314807.3   &  0.076  &   41.87 $\pm$ 0.01    &   1291 $\pm$ 20     &   2049 $\pm$ 35     &   7.38 $\pm$ 0.02     &  7.18  $\pm$ 0.02     &     72 $\pm$ 18           &   16   \\
J111407.35-000031.1   &  0.073  &   41.41 $\pm$ 0.01    &    954 $\pm$ 23     &   1519 $\pm$ 33     &   6.90 $\pm$ 0.02     &  6.70  $\pm$ 0.02     &    125 $\pm$ 10           &   25   \\
J112229.65+214815.5   &  0.061  &   41.44 $\pm$ 0.01    &    994 $\pm$ 16     &   1683 $\pm$ 36     &   6.95 $\pm$ 0.02     &  6.80  $\pm$ 0.02     &    125 $\pm$ 7            &   29   \\
J112229.65+214815.5   &  0.100  &   41.71 $\pm$ 0.01    &   1318 $\pm$ 75     &   2027 $\pm$ 70     &   7.32 $\pm$ 0.05     &  7.09  $\pm$ 0.05     &    176 $\pm$ 20           &   21   \\
J112545.34+240823.9   &  0.024  &   40.20 $\pm$ 0.01    &    688 $\pm$ 16     &   1211 $\pm$ 33     &   6.05 $\pm$ 0.02     &  5.94  $\pm$ 0.02     &     82 $\pm$ 8            &   25   \\
J113101.10+134539.6   &  0.092  &   41.83 $\pm$ 0.01    &   1087 $\pm$ 26     &   1826 $\pm$ 37     &   7.21 $\pm$ 0.02     &  7.06  $\pm$ 0.02     &    171 $\pm$ 14           &   26   \\
J113111.93+100231.3   &  0.074  &   41.25 $\pm$ 0.02    &    930 $\pm$ 33     &   1785 $\pm$ 112    &   6.80 $\pm$ 0.03     &  6.77  $\pm$ 0.03     &    130 $\pm$ 18           &   14   \\
J113913.91+335551.1   &  0.033  &   41.53 $\pm$ 0.01    &    834 $\pm$ 19     &   1394 $\pm$ 48     &   6.84 $\pm$ 0.02     &  6.68  $\pm$ 0.02     &    112 $\pm$ 15           &   32   \\
J115333.22+095408.4   &  0.069  &   41.62 $\pm$ 0.01    &    983 $\pm$ 16     &   1844 $\pm$ 35     &   7.02 $\pm$ 0.02     &  6.97  $\pm$ 0.02     &    130 $\pm$ 11           &   25   \\
J120012.47+183542.9   &  0.066  &   40.92 $\pm$ 0.01    &    862 $\pm$ 47     &   1571 $\pm$ 61     &   6.58 $\pm$ 0.05     &  6.50  $\pm$ 0.05     &    136 $\pm$ 13           &   19   \\
J121157.48+055801.1   &  0.068  &   41.74 $\pm$ 0.01    &   1012 $\pm$ 13     &   1984 $\pm$ 36     &   7.10 $\pm$ 0.01     &  7.09  $\pm$ 0.01     &    119 $\pm$ 12           &   22   \\
J122307.79+192337.0   &  0.076  &   41.33 $\pm$ 0.01    &   1079 $\pm$ 35     &   1832 $\pm$ 101    &   6.97 $\pm$ 0.03     &  6.83  $\pm$ 0.03     &    122 $\pm$ 12           &   21   \\
J123651.17+453904.1   &  0.030  &   41.24 $\pm$ 0.01    &    863 $\pm$ 16     &   1601 $\pm$ 47     &   6.73 $\pm$ 0.02     &  6.67  $\pm$ 0.02     &     97 $\pm$ 7            &   29   \\
J123932.59+342221.3   &  0.084  &   41.53 $\pm$ 0.01    &    898 $\pm$ 56     &   1540 $\pm$ 109    &   6.90 $\pm$ 0.06     &  6.77  $\pm$ 0.06     &     84 $\pm$ 7            &   32   \\
J124319.97+025256.1   &  0.087  &   41.69 $\pm$ 0.01    &    752 $\pm$ 16     &   1276 $\pm$ 29     &   6.81 $\pm$ 0.02     &  6.67  $\pm$ 0.02     &    112 $\pm$ 12           &   26   \\
J130456.96+395529.7   &  0.028  &   40.42 $\pm$ 0.01    &    915 $\pm$ 34     &   1431 $\pm$ 124    &   6.40 $\pm$ 0.03     &  6.19  $\pm$ 0.03     &     92 $\pm$ 6            &   23   \\
J131142.56+331612.7   &  0.078  &   41.29 $\pm$ 0.01    &   1145 $\pm$ 33     &   2086 $\pm$ 41     &   7.01 $\pm$ 0.03     &  6.93  $\pm$ 0.03     &    106 $\pm$ 14           &   16   \\
J131305.81+012755.9   &  0.029  &   40.85 $\pm$ 0.01    &    868 $\pm$ 12     &   1599 $\pm$ 28     &   6.56 $\pm$ 0.01     &  6.49  $\pm$ 0.01     &    108 $\pm$ 5            &   36   \\
J131905.95+310852.7   &  0.032  &   40.97 $\pm$ 0.01    &   1391 $\pm$ 38     &   2063 $\pm$ 61     &   7.03 $\pm$ 0.03     &  6.77  $\pm$ 0.03     &    137 $\pm$ 6            &   38   \\
J134240.09+022524.4   &  0.075  &   41.03 $\pm$ 0.01    &    956 $\pm$ 57     &   1842 $\pm$ 60     &   6.73 $\pm$ 0.05     &  6.70  $\pm$ 0.05     &    105 $\pm$ 14           &   16   \\
J134401.90+255628.3   &  0.062  &   41.33 $\pm$ 0.01    &   1068 $\pm$ 58     &   1651 $\pm$ 43     &   6.96 $\pm$ 0.05     &  6.74  $\pm$ 0.05     &    140 $\pm$ 9            &   25   \\
J140659.58+231738.6   &  0.061  &   40.73 $\pm$ 0.01    &    965 $\pm$ 48     &   1400 $\pm$ 87     &   6.59 $\pm$ 0.05     &  6.31  $\pm$ 0.05     &     97 $\pm$ 8            &   26   \\
J141434.52+293428.2   &  0.076  &   41.29 $\pm$ 0.01    &    844 $\pm$ 29     &   1376 $\pm$ 39     &   6.73 $\pm$ 0.03     &  6.55  $\pm$ 0.03     &     75 $\pm$ 15           &   20   \\
J143658.68+164513.6   &  0.072  &   40.93 $\pm$ 0.01    &    770 $\pm$ 24     &   1418 $\pm$ 55     &   6.49 $\pm$ 0.03     &  6.42  $\pm$ 0.03     &     73 $\pm$ 10           &   17   \\
J143708.46+074013.6   &  0.087  &   41.24 $\pm$ 0.01    &   1089 $\pm$ 45     &   1956 $\pm$ 66     &   6.94 $\pm$ 0.04     &  6.84  $\pm$ 0.04     &     98 $\pm$ 13           &   16   \\
J151356.88+481012.1   &  0.079  &   41.63 $\pm$ 0.01    &    737 $\pm$ 30     &   1270 $\pm$ 55     &   6.77 $\pm$ 0.04     &  6.64  $\pm$ 0.04     &    124 $\pm$ 16           &   21   \\
J152209.56+451124.0   &  0.066  &   41.32 $\pm$ 0.01    &    900 $\pm$ 28     &   1886 $\pm$ 251    &   6.80 $\pm$ 0.03     &  6.85  $\pm$ 0.03     &    128 $\pm$ 12           &   18   \\
J152324.42+551855.3   &  0.039  &   41.12 $\pm$ 0.01    &   1086 $\pm$ 34     &   1717 $\pm$ 91     &   6.88 $\pm$ 0.03     &  6.67  $\pm$ 0.03     &    128 $\pm$ 7            &   33   \\
J152940.58+302909.3   &  0.036  &   41.69 $\pm$ 0.01    &   1073 $\pm$ 22     &   1823 $\pm$ 54     &   7.13 $\pm$ 0.02     &  6.99  $\pm$ 0.02     &    107 $\pm$ 5            &   44   \\
J155640.90+121717.9   &  0.036  &   41.05 $\pm$ 0.01    &   1131 $\pm$ 26     &   2002 $\pm$ 35     &   6.88 $\pm$ 0.02     &  6.78  $\pm$ 0.02     &    149 $\pm$ 9            &   30   \\
J160746.00+345048.9   &  0.054  &   41.53 $\pm$ 0.01    &    749 $\pm$ 7      &   1422 $\pm$ 14     &   6.74 $\pm$ 0.01     &  6.69  $\pm$ 0.01     &     80 $\pm$ 10           &   28   \\
J161527.67+403153.6   &  0.084  &   41.35 $\pm$ 0.01    &    868 $\pm$ 39     &   1608 $\pm$ 54     &   6.79 $\pm$ 0.04     &  6.72  $\pm$ 0.04     &    137 $\pm$ 18           &   17   \\
J161809.36+361957.8   &  0.034  &   41.16 $\pm$ 0.01    &    578 $\pm$ 11     &    896 $\pm$ 27     &   6.34 $\pm$ 0.02     &  6.11  $\pm$ 0.02     &     87 $\pm$ 8            &   30   \\
J161951.31+405847.3   &  0.038  &   41.27 $\pm$ 0.01    &   1020 $\pm$ 15     &   1746 $\pm$ 26     &   6.89 $\pm$ 0.01     &  6.76  $\pm$ 0.01     &    114 $\pm$ 10           &   26   \\
J162930.01+420703.2   &  0.072  &   41.37 $\pm$ 0.01    &    816 $\pm$ 14     &   1440 $\pm$ 36     &   6.74 $\pm$ 0.02     &  6.63  $\pm$ 0.02     &    101 $\pm$ 11           &   22   \\
J163501.46+305412.1   &  0.054  &   41.63 $\pm$ 0.01    &    854 $\pm$ 40     &   1261 $\pm$ 145    &   6.90 $\pm$ 0.04     &  6.63  $\pm$ 0.04     &    130 $\pm$ 14           &   23   \\
J210226.54+000702.3   &  0.052  &   40.76 $\pm$ 0.01    &    806 $\pm$ 45     &   1466 $\pm$ 46     &   6.45 $\pm$ 0.05     &  6.37  $\pm$ 0.05     &     96 $\pm$ 14           &   15   \\
J210533.44+002829.3   &  0.054  &   41.21 $\pm$ 0.01    &    853 $\pm$ 17     &   1429 $\pm$ 27     &   6.71 $\pm$ 0.02     &  6.55  $\pm$ 0.02     &     81 $\pm$ 9            &   23   \\
J010712.03+140844.9   &  0.077  &   41.42 $\pm$ 0.01    &    597 $\pm$ 184    &    998 $\pm$ 170    &   6.48 $\pm$ 0.28     &  6.32  $\pm$ 0.28     &     38 $\pm$ 4$\rm^a$     &   15   \\
J024912.86-081525.7   &  0.030  &   40.21 $\pm$ 0.01    &    542 $\pm$ 19     &    915 $\pm$ 46     &   5.84 $\pm$ 0.03     &  5.69  $\pm$ 0.03     &     53 $\pm$ 3$\rm^a$     &   18   \\
J080629.80+241955.6   &  0.041  &   40.71 $\pm$ 0.01    &    629 $\pm$ 19     &   1067 $\pm$ 39     &   6.20 $\pm$ 0.03     &  6.06  $\pm$ 0.03     &     71 $\pm$ 5$\rm^a$     &   20   \\
J080907.57+441641.4   &  0.054  &   40.90 $\pm$ 0.01    &    692 $\pm$ 27     &   1150 $\pm$ 42     &   6.38 $\pm$ 0.04     &  6.22  $\pm$ 0.04     &     65 $\pm$ 3$\rm^a$     &   21   \\
J081550.23+250640.9   &  0.073  &   40.93 $\pm$ 0.02    &    568 $\pm$ 61     &    895 $\pm$ 90     &   6.21 $\pm$ 0.10     &  6.00  $\pm$ 0.10     &     65 $\pm$ 2$\rm^a$     &   12   \\
J082912.68+500652.3   &  0.044  &   41.28 $\pm$ 0.01    &    597 $\pm$ 7      &   1002 $\pm$ 16     &   6.42 $\pm$ 0.01     &  6.26  $\pm$ 0.01     &     60 $\pm$ 2$\rm^a$     &   29   \\
J094057.19+032401.2   &  0.061  &   41.46 $\pm$ 0.01    &    738 $\pm$ 21     &   1206 $\pm$ 45     &   6.69 $\pm$ 0.03     &  6.51  $\pm$ 0.03     &     82 $\pm$ 3$\rm^a$     &   20   \\
J094529.36+093610.4   &  0.013  &   40.52 $\pm$ 0.01    &    907 $\pm$ 11     &   1767 $\pm$ 27     &   6.44 $\pm$ 0.01     &  6.42  $\pm$ 0.01     &     76 $\pm$ 2$\rm^a$     &   34   \\
J095151.82+060143.6   &  0.093  &   41.00 $\pm$ 0.02    &    742 $\pm$ 100    &   1192 $\pm$ 139    &   6.48 $\pm$ 0.12     &  6.29  $\pm$ 0.12     &     76 $\pm$ 6$\rm^a$     &   11   \\
J101627.33-000714.5   &  0.094  &   41.17 $\pm$ 0.03    &    648 $\pm$ 34     &   1109 $\pm$ 90     &   6.44 $\pm$ 0.05     &  6.31  $\pm$ 0.05     &     55 $\pm$ 7$\rm^a$     &    8   \\
J102348.44+040553.7   &  0.099  &   40.96 $\pm$ 0.02    &    812 $\pm$ 181    &    869 $\pm$ 108    &   6.55 $\pm$ 0.20     &  5.99  $\pm$ 0.20     &     91 $\pm$ 13$\rm^a$    &    9   \\
J111031.61+022043.2   &  0.079  &   41.37 $\pm$ 0.01    &    671 $\pm$ 15     &   1100 $\pm$ 30     &   6.56 $\pm$ 0.02     &  6.39  $\pm$ 0.02     &     77 $\pm$ 3$\rm^a$     &   16   \\
J112526.51+022039.0   &  0.049  &   41.00 $\pm$ 0.01    &    843 $\pm$ 30     &   1305 $\pm$ 48     &   6.60 $\pm$ 0.03     &  6.37  $\pm$ 0.03     &     87 $\pm$ 5$\rm^a$     &   20   \\
J114339.49-024316.3   &  0.094  &   41.32 $\pm$ 0.01    &    746 $\pm$ 40     &   1192 $\pm$ 72     &   6.64 $\pm$ 0.05     &  6.44  $\pm$ 0.05     &     97 $\pm$ 5$\rm^a$     &   22   \\
J121518.23+014751.1   &  0.071  &   41.28 $\pm$ 0.01    &    636 $\pm$ 22     &   1036 $\pm$ 38     &   6.47 $\pm$ 0.03     &  6.29  $\pm$ 0.03     &     81 $\pm$ 3$\rm^a$     &   18   \\
J122342.82+581446.2   &  0.015  &   40.42 $\pm$ 0.01    &    706 $\pm$ 13     &   1049 $\pm$ 32     &   6.17 $\pm$ 0.02     &  5.91  $\pm$ 0.02     &     45 $\pm$ 2$\rm^a$     &   26   \\
J124035.82-002919.4   &  0.081  &   41.76 $\pm$ 0.01    &    728 $\pm$ 11     &   1133 $\pm$ 31     &   6.82 $\pm$ 0.02     &  6.60  $\pm$ 0.02     &     56 $\pm$ 3$\rm^a$     &   19   \\
J125055.28-015556.7   &  0.081  &   41.51 $\pm$ 0.02    &    849 $\pm$ 21     &   1428 $\pm$ 73     &   6.84 $\pm$ 0.02     &  6.69  $\pm$ 0.02     &     66 $\pm$ 4$\rm^a$     &   15   \\
J131926.52+105610.9   &  0.064  &   41.55 $\pm$ 0.01    &    671 $\pm$ 13     &   1040 $\pm$ 31     &   6.65 $\pm$ 0.02     &  6.42  $\pm$ 0.02     &     47 $\pm$ 3$\rm^a$     &   23   \\
J143450.62+033842.5   &  0.028  &   40.27 $\pm$ 0.01    &    708 $\pm$ 34     &   1289 $\pm$ 54     &   6.11 $\pm$ 0.04     &  6.03  $\pm$ 0.04     &     57 $\pm$ 3$\rm^a$     &   22   \\
J144052.60-023506.2   &  0.045  &   41.18 $\pm$ 0.01    &    674 $\pm$ 18     &   1087 $\pm$ 43     &   6.48 $\pm$ 0.03     &  6.29  $\pm$ 0.03     &     73 $\pm$ 8$\rm^a$     &   28   \\
J144705.46+003653.2   &  0.096  &   41.14 $\pm$ 0.02    &    924 $\pm$ 44     &   1495 $\pm$ 56     &   6.75 $\pm$ 0.04     &  6.56  $\pm$ 0.04     &     64 $\pm$ 4$\rm^a$     &    9   \\
J145045.54-014752.9   &  0.099  &   41.62 $\pm$ 0.01    &   1086 $\pm$ 96     &   1690 $\pm$ 250    &   7.11 $\pm$ 0.08     &  6.89  $\pm$ 0.08     &    138 $\pm$ 6$\rm^a$     &   17   \\
J155005.95+091035.7   &  0.092  &   41.73 $\pm$ 0.01    &    572 $\pm$ 37     &    988 $\pm$ 121    &   6.59 $\pm$ 0.06     &  6.46  $\pm$ 0.06     &     78 $\pm$ 6$\rm^a$     &   18   \\
J162636.40+350242.1   &  0.034  &   40.63 $\pm$ 0.01    &    578 $\pm$ 21     &    828 $\pm$ 35     &   6.09 $\pm$ 0.03     &  5.80  $\pm$ 0.03     &     52 $\pm$ 1$\rm^a$     &   24   \\
J163159.59+243740.2   &  0.044  &   41.08 $\pm$ 0.01    &    649 $\pm$ 10     &    958 $\pm$ 20     &   6.40 $\pm$ 0.02     &  6.13  $\pm$ 0.02     &     66 $\pm$ 2$\rm^a$     &   24   \\
J172759.14+542147.0   &  0.100  &   41.28 $\pm$ 0.02    &    668 $\pm$ 39     &   1055 $\pm$ 80     &   6.52 $\pm$ 0.05     &  6.31  $\pm$ 0.05     &     67 $\pm$ 8$\rm^a$     &    8   \\
J205822.14-065004.3   &  0.074  &   41.61 $\pm$ 0.01    &    655 $\pm$ 8      &   1101 $\pm$ 19     &   6.65 $\pm$ 0.01     &  6.50  $\pm$ 0.01     &     58 $\pm$ 3$\rm^a$     &   17   \\
J221139.16-010534.9   &  0.092  &   40.95 $\pm$ 0.02    &    604 $\pm$ 50     &   1104 $\pm$ 64     &   6.28 $\pm$ 0.07     &  6.20  $\pm$ 0.07     &     68 $\pm$ 7$\rm^a$     &   10   \\
J230649.77+005023.3   &  0.061  &   40.93 $\pm$ 0.01    &    851 $\pm$ 46     &   1508 $\pm$ 48     &   6.58 $\pm$ 0.05     &  6.47  $\pm$ 0.05     &     65 $\pm$ 3$\rm^a$     &   16   
\enddata
	\label{tab:NLS1_table}
	\tablenotetext{}{\textbf{Notes.} Column 1: galaxy name; Column 2: redshift;  Column 3: luminosity of ${\rm H\alpha}$; Column 4: line dispersion of ${\rm H\alpha}$; Column 5: FWHM of ${\rm H\alpha}$; Column 6:black hole mass calculated using $\sigma_{\rm H\alpha}$. The errors represent the propagated errors from the uncertainties of the line width and luminosity, without considering systematic errors, e.g., the scatter of the size-luminosity relation, the uncertainty of the virial factor, etc; Column 7: black hole mass calculated using FWHM$_{\rm H\alpha}$; Column 8: stellar velocity dispersion; Column 8: signal-to-noise ratio at 5100\AA\ of the SDSS spectra.}
	\tablenotetext{a}{Stellar velocity dispersions are taken from Xiao et al. (2011).}
	
\end{deluxetable*}

\section{Result}

\subsection{\mbh-$\sigma_{*}$ relation of NLS1s}

\begin{figure*}
\includegraphics[width = 1.0\textwidth]{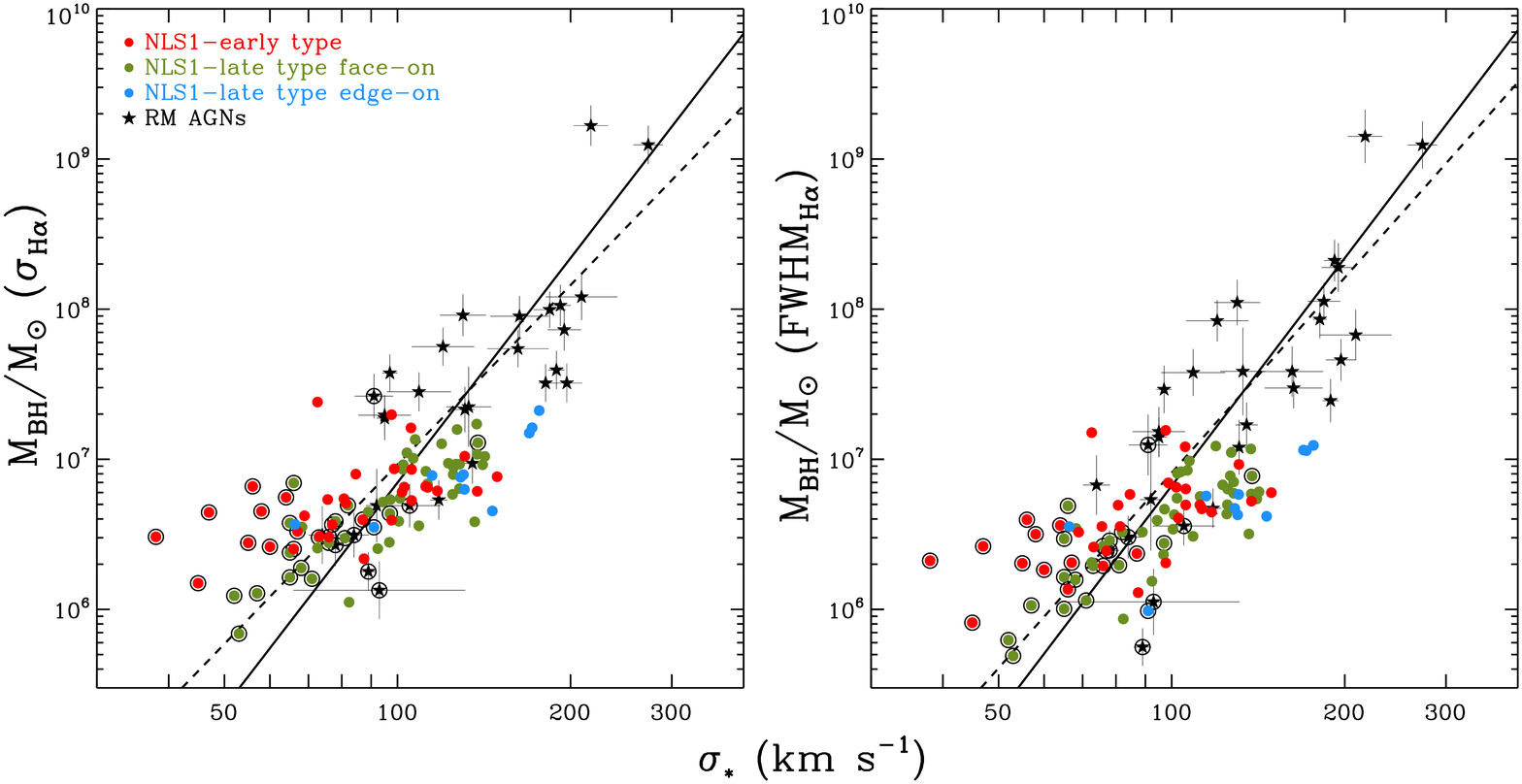}
\caption{\mbh-$\sigma_{*}$ relations of 93 NLS1s with \mbh\ estimated with \sigmaha\ (left) 
and \fwha\ (right), respectively. The morphology and inclination of each galaxy
is represented by different colors: early-type (red), more face-on late-type 
(green), and more edge-on late-type galaxies (blue).  
The solid line represents the best-fit \msigma\ relation of the joint sample of reverberation-mapped AGNs and quiescent galaxies, while the dotted line represents the best-fit \msigma\ relation of the reverberation-mapped AGNs only (see Appendix). 
Six NLS1s among the reverberation-mapped AGNs are denoted with encircled stars. The \sigmastar\ values adopted from \cite{Xiao2011} are represented by
encircled filled circles. 
\label{fig:msigma}}
\end{figure*}

We compare the 93 NLS1 with the RM AGNs and quiescent galaxies in the \msigma\ plane 
in Figure 4. In the left panel, \mbh\ is determined using the line dispersion of the Balmer lines and the virial factor log f = 0.65, while \mbh\ in the right panel is estimated using the FWHM of the Balmer lines and the virial factor log f = 0.05
(see appendix for the determination of the virial factors). 
In general, NLS1s seem to show a consistent \msigma\ relation
compared to the RM AGNs. With respect to the best-fit \msigma\ relation
obtained for the joint sample of the RM AGNs and quiescent galaxies (solid line),
the average offset of the NLS1s is $\Delta log$ \mbh $= 0.04\pm0.06$
in the left panel, and $\Delta log$ \mbh $= -0.08\pm0.06$ in the right panel,
suggesting that NLS1s follow the same \msigma\ relation as other local galaxies.
When we compare NLS1s with the best-fit \msigma\ relation of quiescent galaxies,
we obtained almost the same result since the best-fit \msigma\ relation is almost
identical between quiescent sample and the joint sample of quiescent and RM AGNs
since the quiescent galaxies are dominant in terms of number and dynamical range
\citep[for details, see][]{Woo2013}.

Similarly, when we compare NLS1s with the best-fit \msigma\ relation of the RM AGNs only (dashed line), we obtain a slightly increased offset $\Delta log$ \mbh $= -0.11\pm 0.04$ and $\Delta log $\mbh $= -0.19\pm 0.05$, respectively for $\sigma$-based \mbh\ and $FWHM$-based \mbh. 
The best-fit \msigma\ relation of the RM AGNs suffers from the effect of the limited mass distribution compared to the quiescent galaxy sample. 
The truncation of the mass distribution of the RM AGNs caused a shallower slope of the \msigma\ relation as discussed in detail by Woo et al. 2013. 
In turn, the offset of the NLS1s with respect to 
this shallow \msigma\ slope becomes slightly negative since the NLS1s are mainly located at the low \mbh\ and low stellar velocity dispersion region. 
Considering the small offset and the limited mass distribution, 
NLS1s seem to show a consistent \msigma\ relation compared to the RM AGNs.

Among NLS1s, there is a large scatter with a clear trend with the host galaxy morphology. Compared to the best-fit \msigma\ relation of the joint sample of quiescent galaxies
and RM AGNs, early-type NLS1s show a positive offset 
($\Delta log$ \mbh $= 0.32\pm 0.10$ and $\Delta log$ \mbh $= 0.20\pm 0.10$,
respectively in the left and right panels) 
while late-type galaxies present a negative offset 
($\Delta log$ \mbh $= -0.13\pm 0.06$ and $\Delta log$ \mbh $= -0.25\pm 0.06$,
respectively in the left and right panels in Figure 4). 
The large difference of the offset between early-type and late-type NLS1 galaxies
may stem from the effect of the rotational broadening in the stellar absorption 
lines since single aperture spectra have been used for measuring the stellar
velocity dispersion.  
To test this scenario, we further divide the late-type NLS1 galaxies into
two groups, i.e, edge-on and face-on galaxies (see Section 3.4 for morphology 
classification), and calculated the mean offset. 
Clearly, the edge-on late type galaxies, which are expected to have
larger rotational broadening in the line-of-sight stellar velocity dispersion
measurements, show the largest negative offset 
($\Delta log$ \mbh$ = -0.47\pm 0.15$ and $\Delta log$ \mbh$ = -0.64\pm 0.15$, 
respectively in the left and right panels in Figure 4), 
while the face-on galaxies
do not show a clear offset ($\Delta log$ \mbh$ = -0.05\pm 0.06$ and $\Delta log$ \mbh$ = -0.17\pm 0.06$, respectively in the left and right panels in Figure 4). 
Thus, we suspect that the large scatter of the NLS1s in the \msigma\ plane
and the systematic trend of the offset with galaxy morphology and inclination
are due to the rotational broadening \citep{Xiao2011,Harris2012, Kang2013, Woo2013, Bellovary+14}.

\subsection{offset from the \mbh-$\sigma_{*}$ relation}

\begin{figure}
\includegraphics[width = 0.48\textwidth]{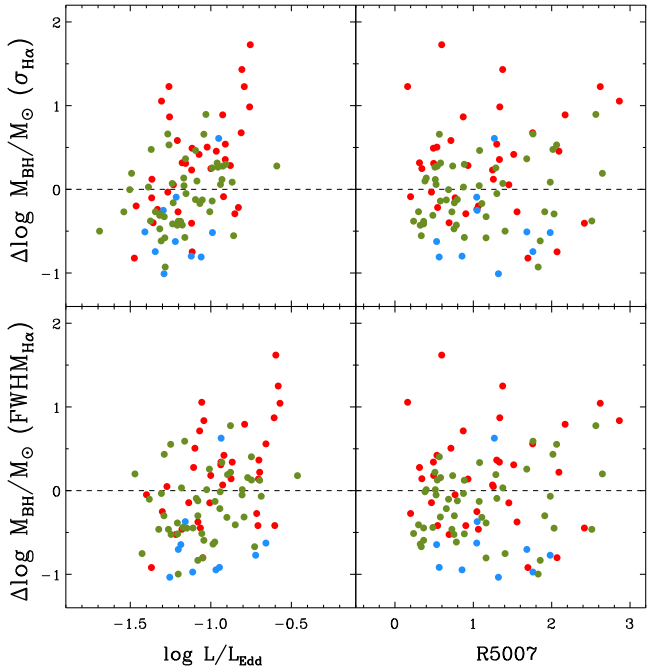}
\caption{Correlations of the offset from the \msigma\ relation with \ledd\ and R5007. 
The colors express same as in Figure \ref{fig:msigma}.
\label{fig:offset}}
\end{figure} 

In this section, we investigate whether the offset from the \msigma\ relation is correlated with 
other AGN parameters, i.e., Eddington ratio, R5007 and R4570.
Note that the offset is calculated with respect to the best-fit \msigma\ relation 
of the joint sample of quiescent galaxies and RM AGNs.
First, we compare the offset with Eddington ratio in Figure~\ref{fig:offset} (left),
finding no significant correlation between the offset and \ledd\ \citep[see consistent results by][]{KomossaXu2007}.
For this comparison, Eddington ratio was determined by dividing the bolometric luminosity by the Eddington luminosity, L$_{\rm Edd}=1.26\times 10^{38} \times M_{\rm BH}$,
using the continuum luminosity at 5100\AA\ as a proxy ($L_{\rm bol}$ = 9\lam\lf) 
\citep{Kaspi2000, Peterson2004}. We also used the H$\alpha$ line luminosity
instead of the continuum luminosity at 5100\AA based on Equation 4, 
and obtained the same results.
Second, we present the comparison between the offset and R5007 in 
Figure~\ref{fig:offset}.
R5007 does not significantly correlate with the offset of NLS1s in general
and in all three different morphology groups. 

\begin{figure}
\includegraphics[width = 0.48\textwidth]{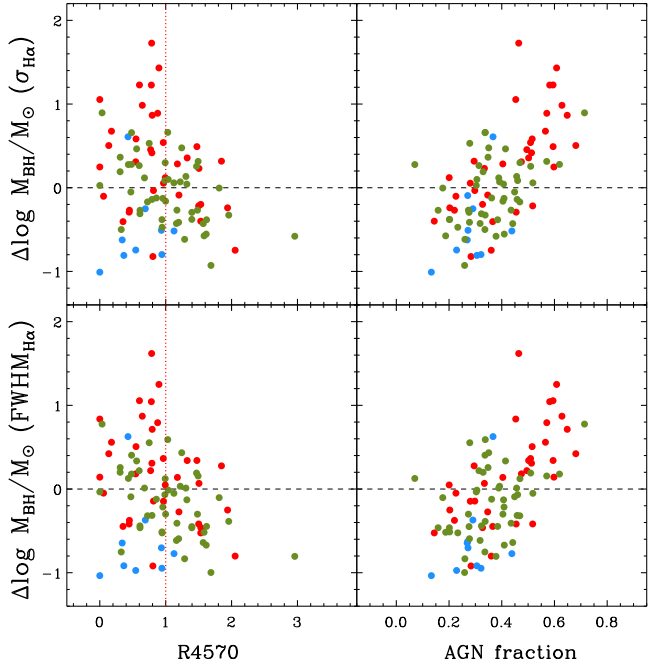}
\caption{Left: comparison of the offset from the AGN \msigma\ relation with R4570 (left)
and AGN fraction (right). 
The colors express same as in Figure \ref{fig:msigma}.
\label{fig:offset_R4570}}
\end{figure}

In Figure~ \ref{fig:offset_R4570}, we compare the \FeII\ strength (R4570)
with the offset from the \msigma\ relation. 
There is a weak correlation between them: while the weak \FeII\ emitters
show both positive and negative offsets, the strong \FeII\ emitters
mainly show negative offset \citep[see the reference line at R4570 $=1$;][]
{Lawrence1988, Zhou2006}. This result implies that NLS1s with strong \FeII\ 
more significantly deviate from the \msigma\ relation. 
The correlation is slightly different for different morphology groups. 
However, the sample size in each morphology group is too small to definitely claim 
any difference.

In addition, we tested whether the offset from the \msigma\ relation is
related with the AGN fraction (see Figure~\ref{fig:offset_R4570} right panels), 
using the AGN fraction determined from the monochromatic flux ratio of AGN to host 
galaxy at 5100\AA. 
We find a good correlation of the offset with the AGN fraction:
the offset positively increases with increasing AGN fraction.
This correlation is also present in each morphology group while
early-type galaxies have on average higher AGN fraction than late type galaxies.
The interpretation of this correlation is not straightforward since
a strong selection effect is involved in measuring stellar velocity dispersion.
For example, if the AGN fraction is higher, then it is more difficult to measure
stellar velocity dispersion. Hence, only massive early-type galaxies
are available at high AGN fraction regime, while most late-type galaxies
hosting high luminosity AGNs are missing from the distribution.

\section{Discussion}

\subsection{The effect of rotational broadening}

The contribution of the rotation of stellar disks can bias stellar velocity 
dispersion measurements due to the rotational broadening of the
stellar absorption lines. For example, if a single-aperture spectrum, which is 
integrated over a large portion of a stellar disk, is used for measuring the second
moment of the absorption lines, 
the rotational effect can significantly increase the velocity dispersion measurements.          
For early-type galaxies the rotation effect is relatively small
since the velocity dispersion is typically higher than rotation velocity.
For example, \citet{Kang2013} reported that the stellar velocity dispersion 
changes by $\sim$10\% as a function of the aperture size, based on the spatially
resolved measurements of 31 early-type galaxies in the \msigma\ sample. 
In contrast, we expect the rotation effect can be substantially larger in late-type
galaxies than in early-type galaxies due to much higher velocity-to-dispersion (V/$\sigma$)
ratios. For disk-dominant late-type galaxies, the inclination to the line-of-sight 
can also play a significant role due to the project effect. Based on 
the n-body smoothed particle hydrodynamic simulations, \cite{Bellovary+14}
reported that bulge stellar velocity dispersion measurements can change by 30\% 
depending on the galaxy inclination.

Since most of the \sigmastar\ measurements for AGN host galaxies are based single-aperture
spectra, the effect of rotation and inclination can play a role in comparing BL AGNs
with NLS1s. Using a sample of low \mbh\ AGN sample, \cite{Xiao2011} showed a clear dependency
of galaxy inclination on the offset from the \msigma\ relation, i.e.,
more inclined galaxies tend to have higher \sigmastar\ and negatively offset,
while more face-on galaxies tend to have lower \sigmastar\ and positively offset.
The observed trend of the offset from the \msigma\ relation with galaxy inclination
in our study is similar to \cite{Xiao2011}, reflecting the same effect of the rotation and inclination of late-type galaxies. 
Thus, the conclusion that NLS1s follow the same \msigma\ relation as BL AGNs
is still limited by the lack of spatially resolved measurements.
To better understand the effect of rotation and inclination,
spatially resolved measurements are required for NLS1s,
which is beyond the scope of the current study.

\subsection{NLS1s versus BL AGNs}

Two different scenarios have been suggested for the evolution of NLS1s.
On the one hand, NLS1s are considered as the precursors of BL AGNs, evolving into BL AGNs.
The low \mbh\ and the high Eddington ratio of NLS1s may imply
that NLS1s are young phase of AGNs \citep{Veron2001, Mathur2001, Boroson2002}. 
On the other hand, NLS1s are viewed as an extension of BL AGNs at the low-mass scale
\citep{McHardy2006}. 
If the high Eddington ratio of NLS1s represents a relatively short-lived accretion phase, 
and the Eddington ratio before and after the strong accretion phase is relatively low,  
then the black hole growth in NLS1 may be insignificant. 
We find no significant evidence that NLS1 offset from the \msigma\ relation of
active and inactive galaxies, suggesting that NLS1s and BL AGNs are similar in
terms of the current black hole growth.
In the case of the host galaxies of NLS1s, there seems no strong difference 
between the environments of NLS1s and BL AGNs \citep{Krongold2001}. 
Also, NLS1s are not preferentially hosted by merging galaxies
\citep{Ryan2007}. Thus, the growth of black holes and host galaxies 
seem to be similar between NLS1 and BL AGNs.

Based on the estimates of the bolometric luminosity of the NLS1s in our sample,
we calculated the mass accretion rate in order to investigate the black hole
growth time scale. For given the range of bolometric luminosity of 10$^{43}$ - 10$^{44}$ \ergs,
we estimate the mass accretion rate as $\sim$0.002 - $\sim$0.02 \msun\ year$^{-1}$.
Thus, in order to accrete a million solar mass to a black hole with a constant 
mass accretion rate of $\sim$0.002 - $\sim$0.02 \msun\ year$^{-1}$, 
it would take 10$^{8}$-10$^{9}$ yrs. The mean Eddington ratio of the NLS1s 
in our sample is $\sim$10\%, for which the e-folding growth time scale is
4$\times$$10^{8}$ yrs. Thus, unless the life time of the AGN activity is 
comparable to this growth time scale, NLS1s are not expected to move up
to the larger \mbh\ direction in the \msigma\ plane \citep[see also discussion by][]
{KomossaXu2007}. 

\subsection{Inclination angle}

NLS1s are often considered as more inclined (pole-on) systems to the line-of-sight 
than BL AGNs, implying that the measured line-of-sight velocity dispersion (line width)
of broad emission lines is relatively narrow due to the projection effect.
If this is the case, then the \mbh\ of NLS1s are significantly underestimated and 
their Eddington ratios are accordingly overestimated. 
However, although there are some evidences that NLS1s are close to pole-on systems 
\citep[e.g.,][]{Fischer2014, Foschini2014}, 
the inclination effect cannot explain the entire NLS1 population
(see discussion by Peterson 2011).  
The implication of the potential inclination effect is that the NLS1s in our sample would 
positively offset toward the high \mbh\ direction, if the black hole masses 
were were estimated after correcting for the velocity projection effect. 
In this scenario, it is difficult to understand why NLS1s have higher black 
hole to galaxy mass ratios compared to BL AGNs and quiescent galaxies. 

We note that 6 NLS1s are included in the sample of the reverberation-mapped AGNs, 
which are used for deriving the average virial factor for type 1 AGNs 
(see Figure 5). The location of the NLS1s in the \msigma\ plane is not
different from that of BL AGNs, implying that the virial factor and inclination
angle of the NLS1s may not be very different from those of BL AGNs,
although the number of NLS1s in the reverberation-mapped AGN sample 
is still small to make a firm conclusion.

\section{Summary \& Conclusion}

We investigated the \msigma\ relation of the present-day NLS1, using 
directly measured stellar velocity dispersions for a sample of 93 NLS1s at z$<$0.1 
selected from the SDSS. We summarize the main results.

\smallskip 
$\bullet$ Compared to the \msigma\ relation derived from the joint sample
of the reverberation-mapped AGNs and inactive galaxies, 
the NLS1s in our sample show no significant offset, 
suggesting that NLS1s are an extension of BL AGNs at lower mass scale.

$\bullet$ Among NLS1s, there is a systematic trend with galaxy inclination, i.e., 
more inclined galaxies have larger \sigmastar\ at fixed \mbh, probably due to 
the contribution of the rotational broadening in the stellar absorption lines. 

$\bullet$ By jointly fitting the \msigma\ relation using the most updated 
reverberation-mapped AGNs and quiescent galaxies,
we obtained the virial factor log f = 0.65 $\pm$ 0.12 (i.e., f = 4.47) and log f = 0.05 $\pm$ 0.12 (i.e., f = 1.12), 
respectively for \mbh\ estimators based on the \sigmahb\ and \fwhb.

\acknowledgements

We thank the anonymous referee for valuable comments, which improved the clarity of the manuscript. 
This work was supported by the National Research Foundation of Korea (NRF) grant funded by the Korea government (MEST; No. 2012-006087). J.H.W acknowledges the support by the Korea Astronomy and Space Science Institute (KASI) grant funded by the Korea government (MEST).

\appendix

\section{A1. The Virial Factor}

The virial factor f in Equation (1) is difficult to determine for individual objects due to the unknown geometry and distribution of the BLR gas \citep[c.f.,][]{Brewer2011, Pancoast2012, Pancoast2014}. Instead, an average $f$ has been determined by scaling the reverberation-mapped AGNs to quiescent galaxies in the \msigma\ plane, assuming that AGN and non-AGN galaxies follow the same \msigma\ relation \citep{Onken2004, Woo2010, Woo2013, Park2012}.
While most of these calibrations have been performed using the virial product ($V^2 \times$ R$_{\rm BLR}$ /G) based on \sigmahb\ as the velocity proxy of the broad-line gas,
a number of black hole mass studies used \fwhb\ for estimating single-epoch \mbh\ because
of the difficulty of measuring \sigmahb\ due to the low S/N of available spectra (e.g., SDSS).
In this case, \fwhb\ is converted to \sigmahb\ with a constant
FWHM/$\sigma$ ratio. However, the FWHM/$\sigma$ ratio has a wide range
since the line profile of the \Hb\ line is not universal \citep{Peterson2004, Collin2006},
hence, a systematic uncertainty is added to the mass estimates.
Here we provide the f factor for \sigmahb-based and \fwhb-based virial products, respectively, by fitting the \msigma\ relation.

For the reverberation-mapped AGNs, we collected and updated the time-lag \citep[e.g.,][]{Zu2011}, \fwhb\ and \sigmahb\ 
as well as stellar velocity dispersion measurements from the literature 
for a sample of 29 AGNs, as listed in Table A1 (see a recent compilation by Woo et al. 2013 and the addition of Grier et al. 2013 and Bentz et al. 2014), after excluding two objects,
PG 1229+204 and PG 1617+175 since their stellar velocity dispersion measurements
are very uncertain (see for example Figure 3 in Grier et al. 2013). 
The \fwhb\ and \sigmahb\ are measured from the rms spectra of each object except for 
the 4th entry of Mrk 817 (see Table A1). 
When there are multiple measurements available for given objects, we calculated the mean of the virial products. 
Note that we often found typos of the quoted values of the time lag
and the H$\beta$ velocity in the literature. 
Thus, we included the reference of the original measurements.

In the case of the quiescent galaxy sample, we used 84 galaxies from the compilation
of Kormendy \& Ho (2013), after excluding 3 galaxies, NGC 2778, NGC 3945, NGC 4382
due to the lack of the lower limit of the black hole mass. 
Note that the choice of the quiescent galaxy sample does not significantly 
change the results presented for the NLS1s since the virial factor is
determined based on the best-fit \msigma\ relation and the \mbh\ of the NLS1s
scales accordingly. A careful comparison of the \msigma\ relation based on
various subsamples of the quiescent galaxies will be presented by Woo et al. 
(in preparation). 

We performed a joint-fit analysis for the combined sample of reverberation-mapped AGNs and quiescent galaxies in order to determine the slope, intercept, and the virial factor, following the joint-fit method as described in Woo et al. 2013: 
\begin{equation}
 \chi^2 = \sum_{i=1}^N \frac{ \left( \mu_i -\alpha -\beta s_i \right)^2}{\sigma_{\mu,i}^2 + 
 \beta^2 \sigma_{s,i}^2 + \epsilon_0^2}  + \sum_{j=1}^M  \frac{ \left( \mu_{{\rm VP},j}  +\log f-\alpha -\beta s_j \right)^2}{\sigma_{\mu_{VP},j}^2 + 
 \beta^2 \sigma_{s,j}^2 + \epsilon_0^2}\  ,
\end{equation}
where $\mu=\log\ $(\mbh/\msun) of quiescent galaxies, 
$\mu_{VP} = \log\ $($V^2 R_{\rm BLR} / G$) of reverberation-mapped AGNs, 
and $s=\log\ $(\sigmastar / 200 \kms),
while $\sigma_{\mu}$, $\sigma_{\mu_{VP}}$, and $\sigma_{s}$
are the measurements uncertainties in $\mu$, $\mu_{VP}$, and $s$, respectively,
and $\epsilon_0$ is intrinsic scatter, which we change for the reduced $\chi^2$ to be unity. 
In Figure A1, we present the best-fit \msigma\ relation for the combined sample.
When \sigmahb\ is used as V in Eq. 1, we obtained  
the intercept $\alpha$ = 8.34 $\pm$ 0.05, the slope $\beta$ = 4.97 $\pm$ 0.28,
and log f = 0.65 $\pm$ 0.12. In the case of \fwhb, we derived $\alpha$ = 8.34 $\pm$ 0.05, $\beta$ = 5.04 $\pm$ 0.28, and log f = 0.05 $\pm$ 0.12. 
The intrinsic scatter of the combined sample is 0.43 $\pm$ 0.03 and 0.43 $\pm$ 0.03, respectively for \sigmahb-based mass and \fwhb-based mass.
The derived f factor and the \msigma\ relation based on the updates of the reverberation and stellar velocity dispersion measurements are consistent with those derived by \cite{Woo2013}. In the case of the \fwhb-based \mbh, 
the best-fit virial factor f = 1.12 is consistent with the value derived by \citet{Collin2006}. 
For future \mbh\ studies, we recommend to use log f = 0.65 $\pm$ 0.12 for the \sigmahb-based \mbh\ estimates, and log f = 0.05 $\pm$ 0.12 for the \fwhb-based \mbh\ estimates.
The derived virial factor is consistent with that determined from the dynamical modeling
based on the velocity-resolved measurements of five AGNs (Pancoast et al. 2014),
which are log f = 0.68 $\pm$ 0.40 and log f = -0.07 $\pm$ 0.40, respectively for
the \sigmahb-based and \fwhb-based black hole masses. 

Note that we did not attempt to use a different \msigma\ relation for pseudo-bulge
galaxies since the \msigma\ relation of the pseudo-bulge galaxies is not well
defined due to the limited dynamical range (see Figure A1). It is not clear
whether pseudo-bulge galaxies offset from the \msigma\ relation of classical bulges
in Figure A1 (see also Bennert et al. 2014). More detailed comparison of pseudo-bulge
galaxies in the \msigma\ plane will be provided by Woo et al. (in preparation)
based on the new measurements from the spatially-resolved kinematics of 9 pseudo-bulge galaxies.  
Thus, in this study we simply combine classical and pseudo bulges in determining the best fit \msigma\ relation. In Figure A1, we used open symbols for pseudo bulge galaxies
following the classification from Kormendy \& Ho 2014 and Ho \& Kim 2014. 

As a consistency check, we fit the \msigma\ relation for the AGN sample only by minimizing
\begin{equation}
 \chi^2 = \sum_{i=1}^N \frac{ \left( \mu_i -\alpha -\beta s_i \right)^2}{\sigma_{\mu,i}^2 + 
 \beta^2 \sigma_{s,i}^2 + \epsilon_0^2}   \  ,
\end{equation}
where we used log f = 0.65 for the \sigmahb-based \mbh\ estimates, and log f = 0.05 for \fwhb-based \mbh. 
Using the \sigmahb-based \mbh, we obtained the best-fit $\alpha$ = 8.16 $\pm$ 0.18, $\beta$ = 3.97 $\pm$ 0.56,
and the intrinsic scatter $\epsilon$ = 0.41 $\pm$ 0.05. 
In the case of the \fwhb-based \mbh, we derived $\alpha$ = 8.21 $\pm$ 0.18, $\beta$ = 4.32 $\pm$ 0.59, 
and $\epsilon$ = 0.43 $\pm$ 0.05. These slopes are consistent with the best-fit slope of the
combined sample within the uncertainties.  
We note that the slope $\alpha$ of the AGN \msigma\ relation does not depend on
the choice of the virial factor in Equation A2.

We emphasize that in our study the \msigma\ relation of the reverberation-mapped AGNs is 
derived with a consistent method adopted for the quiescent galaxies (see Park et al. 2012), 
while other studies of AGN \msigma\ relation often utilized somewhat different method,
without including an iterative fitting process with intrinsic scatter.
Compared to Grier et al. (2013), for example, we obtained a different \msigma\ relation, 
hence, the virial factor even if we used the compiled values 
in their table. This discrepancy seems to stem from the treatment of the intrinsic scatter
since we obtained the same results as Grier et al. (2013) when we excluded the intrinsic
scatter in the fitting process.

\begin{figure}
\includegraphics[width = \textwidth]{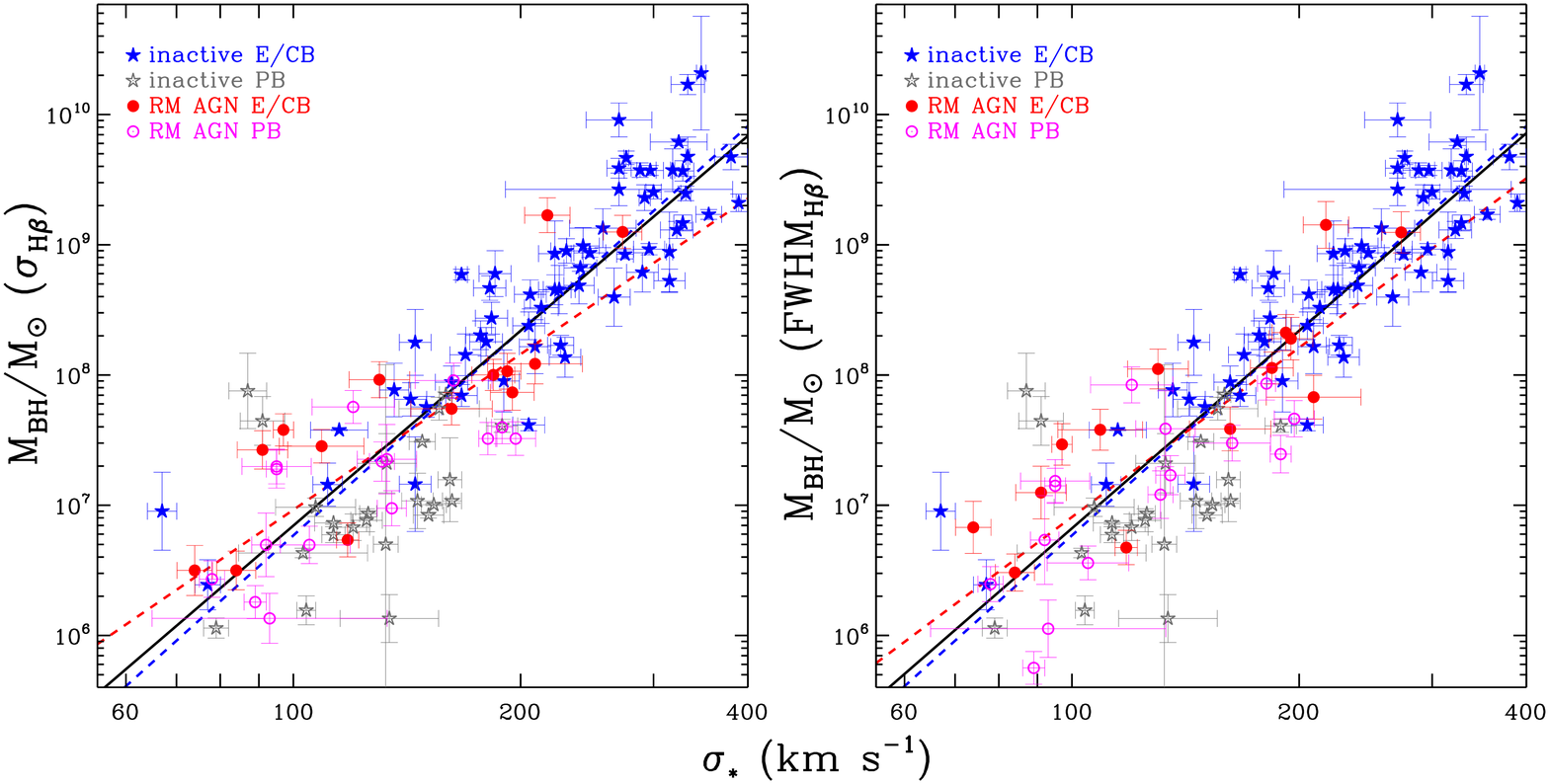}
\caption{\msigma\ relation of quiescent (blue and grey stars) and
active (red and magenta circles) galaxies with AGN \mbh estimated from \sigmahb\ (left) 
and \fwhb\ (right), respectively .
Based on the joint fit result, we used $log f=0.65\pm0.12$ for the \sigmahb-based \mbh\ 
and $log f=0.05\pm0.12$ for the \fwhb-based \mbh. 
The solid line represents the best fit for the combined sample of quiescent galaxies and the reverberation-mapped AGNs while the red dashed line represents the best-fit
for the reverberation-mapped AGN only. We also present the best-fit \msigma\ relation
for the quiescent galaxy sample only (blue dashed line), which is consistent with the
best fit of the joint sample. This is due to the fact that the quiescent galaxy sample 
has a similar dynamical range compared to the combined sample. 
Psuedo-bulge galaxies are denoted with open symbols
(magenta circles for active galaxies; grey stars for quiescent galaxies) while
ellipticals and pseudo-bulges are represented by filled symbols.
}
\label{fig:jointfit}
\end{figure}

\clearpage

\begin{deluxetable*}{lcccc ccccc c}
	\tablecolumns{11}
	\tablewidth{0pc}
	\tablecaption{Virial Products and \sigmastar\ of the Reverberation-mapped AGNs}
	\tablehead{
		\colhead{Name}&
		\colhead{}&
		\colhead{$\tau_{\rm H\beta}$}&
		\colhead{Ref.}&
		\colhead{$\sigma_{\rm line}$}&
		\colhead{$\rm FWHM_{\rm H\beta}$}&
		\colhead{Ref.}&
		\colhead{$\rm VP(\sigma_{\rm line})$}&
		\colhead{$\rm VP(FWHM_{\rm H\beta})$}&
		\colhead{$\sigma_{*}$}&
		\colhead{Ref.}
		\\
		\colhead{}&
		\colhead{}&
		\colhead{(days)}&
		\colhead{}&
		\colhead{($\rm km\ s^{-1}$)}&
		\colhead{($\rm km\ s^{-1}$)}&
		\colhead{}&
		\colhead{($10^{6}\ \rm M_{\odot}$)}&
		\colhead{($10^{6}\ \rm M_{\odot}$)}&
		\colhead{($\rm km\ s^{-1}$)}&
		\colhead{}
		\\
		\colhead{(1)}&
		\colhead{}&
		\colhead{(2)}&
		\colhead{(3)}&
		\colhead{(4)}&
		\colhead{(5)}&
		\colhead{(6)}&
		\colhead{(7)}&
		\colhead{(8)}&
		\colhead{(9)}&
		\colhead{(10)}
	}
	\startdata
	3C 120         &      	 & 	$	27.2   ^{	+1.1  }_{  -1.1 }	$	&  1 	&   1514  $\pm$   65 	&  2539  $\pm$   466  &   1	&  $   12.2  ^{  0.9  }_{    -0.9  }	$   	&  $     34.2  ^{    9.0 }_{    -9.0  }	$	&   162  $\pm$   20	&  13 \\
	3C 390.3       &      	 & 	$	47.9   ^{	+2.4  }_{  -4.2 }	$	&  2 	&   5455  $\pm$  278 	& 10872  $\pm$  1670  &   2	&  $  278.1  ^{ 24.4  }_{   -31.6  }	$   	&  $   1104.8  ^{  246.3 }_{  -258.8  }	$	&   273  $\pm$   16	&  14 \\
	Ark 120        &    		 & 	$	35.7   ^{	+6.7  }_{  -9.2 }	$	&  3 	&   1959  $\pm$  109 	&  5536  $\pm$   297  &   4	&  $   26.7  ^{  5.4  }_{    -7.2  }	$   	&  $    213.5  ^{   43.2 }_{   -57.4  }	$	&  					 	&  	\\
	&    		 & 	$	29.7   ^{	+3.3  }_{  -5.9 }	$	&  3 	&   1884  $\pm$   48 	&  5284  $\pm$   203  &   4	&  $   20.6  ^{  2.4  }_{    -4.2  }	$   	&  $    161.8  ^{   20.0 }_{   -33.3  }	$	&   					 	&   	\\
	&	mean   &  													&		&				     		  	&  		       	 	 &			& 	$ 	 23.7  ^{  3.0  }_{    -4.2  }	$ 		& 	$ 	  187.7  ^{   23.8 }_{   -33.2  }	$	&   192  $\pm$    8	&  15	\\
	\\                                                                                                                                                                                                                                         
	Arp 151        &       	 & 	$	3.6    ^{	+0.7  }_{  -0.2 }	$	&  5 	&   1295  $\pm$   37 	&  2458  $\pm$    82  &   6	&  $    1.2  ^{  0.2  }_{    -0.1  }	$   	&  $      4.2  ^{    0.8 }_{    -0.3  }	$	&   118  $\pm$    4	&  16 \\
	Mrk 50         &      	 & 	$	10.6   ^{	+0.8  }_{  -0.9 }	$	&  7 	&   1740  $\pm$  101 	&  4039  $\pm$   606$\rm^a$  &   7	&  $    6.3  ^{  0.7  }_{    -0.7  }	$   	&  $     33.7  ^{    7.6 }_{    -7.7  }	$	&   109  $\pm$   14	&   7	\\
	Mrk 79	      &    		 & 	$	25.5   ^{	+2.9  }_{ -14.4 }	$	&  3 	&   2137  $\pm$  375 	&  5086  $\pm$  1436  &   4	&  $   22.7  ^{  6.2  }_{   -14.0  }	$   	&  $    128.7  ^{   53.4 }_{   -89.0  }	$	&    					 	& 	 	\\
	&    		 & 	$	30.9   ^{	+1.4  }_{  -2.1 }	$	&  3 	&   1683  $\pm$   72 	&  4219  $\pm$   262  &   4	&  $   17.1  ^{  1.3  }_{    -1.6  }	$   	&  $    107.3  ^{   10.6 }_{   -11.9  }	$	&    					 	& 	 	\\
	&    		 & 	$	17.2   ^{	+7.3  }_{  -2.2 }	$	&  3 	&   1854  $\pm$   72 	&  5251  $\pm$   533  &   4	&  $   11.5  ^{  4.9  }_{    -1.6  }	$   	&  $     92.5  ^{   41.5 }_{   -17.8  }	$	&   					 	&   	\\
	&    		 & 	$	43.6   ^{	+1.7  }_{  -0.8 }	$	&  3 	&   1883  $\pm$  246 	&  2786  $\pm$   390  &   4	&  $   30.2  ^{  5.7  }_{    -5.6  }	$   	&  $     66.0  ^{   13.3 }_{   -13.1  }	$	&   					 	&   	\\
	&	mean	 & 													&		&				     		  	&     	       	 	 &			& 	$ 	 20.4  ^{  2.5  }_{	  -3.8  }	$ 		& 	$ 		98.7  ^{   17.4 }_{   -23.1  }	$	&   130  $\pm$   12	&  14 \\
	\\                                                                                                                                                                                                                                         
	Mrk 110        &    		 & 	$	25.3   ^{	+2.3  }_{ -13.1 }	$	&  3 	&   1196  $\pm$  141 	&  1494  $\pm$   802  &   4	&  $    7.1  ^{  1.3  }_{    -3.8  }	$   	&  $     11.0  ^{    8.4 }_{   -10.1  }	$	&    					 	&   	\\
	&    		 & 	$	33.9   ^{	+6.1  }_{  -5.3 }	$	&  3 	&   1115  $\pm$  103 	&  1381  $\pm$   528  &   4	&  $    8.2  ^{  1.8  }_{    -1.7  }	$   	&  $     12.6  ^{    7.2 }_{    -7.1  }	$	&    					 	&   	\\
	&    		 & 	$	21.5   ^{	+2.2  }_{  -2.1 }	$	&  3 	&    755  $\pm$   29 	&  1521  $\pm$    59  &   4	&  $    2.4  ^{  0.3  }_{    -0.3  }	$   	&  $      9.7  ^{    1.1 }_{    -1.1  }	$	&    					 	&   	\\
	&  mean 	 & 						 							&		&				     		  	&     	       	 	 &			& 	$ 	  5.9  ^{  0.8  }_{	  -1.4  }	$ 		& 	$ 		11.1  ^{    3.7 }_{    -4.1  }	$	&    91  $\pm$    7	&  17 \\
	\\                                                                                                                                                                                                                                         
	Mrk 202        &       	 & 	$	3.5    ^{	+0.1  }_{  -0.1 }	$	&  5 	&    962  $\pm$   67 	&  1794  $\pm$   181  &   6	&  $    0.6  ^{  0.1  }_{    -0.1  }	$   	&  $      2.2  ^{    0.3 }_{    -0.3  }	$	&    78  $\pm$    3	&  16 \\
	Mrk 279        &      	 & 	$	18.3   ^{	+1.2  }_{  -1.1 }	$	&  3 	&   1420  $\pm$   96 	&  3385  $\pm$   349  &   4	&  $    7.2  ^{  0.8  }_{    -0.8  }	$   	&  $     40.9  ^{    6.5 }_{    -6.5  }	$	&   197  $\pm$   12	&  14 \\
	Mrk 509        &     	 & 	$	69.9   ^{	+0.3  }_{  -0.3 }	$	&  3 	&   1276  $\pm$   28 	&  2715  $\pm$   101  &   4	&  $   22.2  ^{  0.7  }_{    -0.7  }	$   	&  $    100.5  ^{    5.3 }_{    -5.3  }	$	&   184  $\pm$   12	&   5	\\
	Mrk 590        &    		 & 	$	19.0   ^{	+1.8  }_{  -2.6 }	$	&  3 	&    789  $\pm$   74 	&  1675  $\pm$   587  &   4	&  $    2.3  ^{  0.4  }_{    -0.4  }	$   	&  $     10.4  ^{    5.2 }_{    -5.3  }	$	&  					 	&   	\\
	&    		 & 	$	19.5   ^{	+2.0  }_{  -4.0 }	$	&  3 	&   1935  $\pm$   52 	&  2566  $\pm$   106  &   4	&  $   14.2  ^{  1.6  }_{    -3.0  }	$   	&  $     25.1  ^{    3.0 }_{    -5.3  }	$	&   					 	&   	\\
	&    		 & 	$	32.6   ^{	+3.5  }_{  -8.8 }	$	&  3 	&   1251  $\pm$   72 	&  2115  $\pm$   575  &   4	&  $   10.0  ^{  1.3  }_{    -2.8  }	$   	&  $     28.5  ^{   11.4 }_{   -13.4  }	$	&   					 	&   	\\
	&    		 & 	$	30.9   ^{	+2.5  }_{  -2.4 }	$	&  3 	&   1201  $\pm$  130 	&  1979  $\pm$   386  &   4	&  $    8.7  ^{  1.5  }_{    -1.5  }	$   	&  $     23.6  ^{    6.8 }_{    -6.8  }	$	&   					 	&   	\\
	&	mean   & 		     			 							&		&				     		  	&     	       	 	 &			& 	$ 	  8.8	 ^{  0.6  }_{    -1.1  }	$ 		& 	$ 		21.9  ^{    3.6 }_{    -4.2  }	$	&   189  $\pm$    6	&  14 \\
	\\                                                                                                                                                                                                                                         
	Mrk 817        &    		 & 	$	20.9   ^{	+2.3  }_{  -2.3 }	$	&  3 	&   1392  $\pm$   78 	&  3515  $\pm$   393  &   4	&  $    7.9  ^{  1.1  }_{    -1.1  }	$   	&  $     50.4  ^{    9.7 }_{    -9.7  }	$	&   					 	&   	\\
	&    		 & 	$	17.2   ^{	+1.9  }_{  -2.7 }	$	&  3 	&   1971  $\pm$   96 	&  4952  $\pm$   537  &   4	&  $   13.0  ^{  1.7  }_{    -2.2  }	$   	&  $     82.3  ^{   15.6 }_{   -18.1  }	$	&   					 	&   	\\
	&    		 & 	$	35.9   ^{	+4.8  }_{  -5.8 }	$	&  3 	&   1729  $\pm$  158 	&  3752  $\pm$   995  &   4	&  $   20.9  ^{  3.9  }_{    -4.3  }	$   	&  $     98.6  ^{   39.3 }_{   -40.3  }	$	&   					 	&   	\\
	&    		 & 	$	10.8   ^{	+1.5  }_{  -1.0 }	$	&  3 	&   2025  $\pm$    5$^b$ 	&  5627  $\pm$    30$^b$  &   8	&  $    8.6  ^{  1.2  }_{    -0.8  }	$   	&  $     66.7  ^{    9.3 }_{    -6.2  }	$	&   					 	&   	\\
	&	mean   & 		    	 		 							&		&				     		  	&     	       	 	 &			& 	$ 	 12.6  ^{  1.1  }_{    -1.3  }	$ 		& 	$ 		74.5  ^{   11.1 }_{   -11.4  }	$	&   120  $\pm$   15	&  14 \\
	\\                                                                                                                                                                                                                                         
	Mrk 1310       &       	 & 	$	4.2    ^{	+0.9  }_{  -0.1 }	$	&  5 	&    921  $\pm$  135 	&  1823  $\pm$   157  &   6	&  $    0.7  ^{  0.2  }_{    -0.1  }	$   	&  $      2.7  ^{    0.7 }_{    -0.3  }	$	&    84  $\pm$    5	&  16 \\
	NGC 3227       &    		 & 	$	10.6   ^{	+6.1  }_{  -6.1 }	$	&  3 	&   2018  $\pm$  174 	&  5278  $\pm$  1117  &   4	&  $    8.4  ^{  5.0  }_{    -5.0  }	$   	&  $     57.6  ^{   37.4 }_{   -37.4  }	$	&  					 	&  	\\
	&			 & 	$	4.4    ^{	+0.3  }_{  -0.5 }	$	&  3 	&   1376  $\pm$   44 	&  3578  $\pm$    83  &   8	&  $    1.6  ^{  0.1  }_{    -0.2  }	$   	&  $     11.0  ^{    0.8 }_{    -1.3  }	$	&   					 	&   	\\
	&	mean   & 		     	 		 							&		&				     		  	&     	       	 	 &			& 	$	  5.0	 ^{  2.5  }_{    -2.5  }	$ 		& 	$ 		34.3  ^{   18.7 }_{   -18.7  }	$	&   133  $\pm$   12	&  18 \\
	\\                                                                                                                                                                                                                                         
	NGC 3516       &      	 & 	$	14.6   ^{	+1.4  }_{  -1.1 }	$	&  3 	&   1591  $\pm$   10 	&  5175  $\pm$    96  &   8	&  $    7.2  ^{  0.7  }_{    -0.5  }	$   	&  $     76.3  ^{    7.6 }_{    -6.1  }	$	&   181  $\pm$    5	&  14 \\
	NGC 3783       &  	  	 & 	$	7.3    ^{	+0.3  }_{  -0.7 }	$	&  3 	&   1753  $\pm$  141 	&  3093  $\pm$   529  &   4	&  $    4.4  ^{  0.5  }_{    -0.7  }	$   	&  $     13.6  ^{    3.3 }_{    -3.5  }	$	&    95  $\pm$   10	&  19 \\
	NGC 4051       &    		 & 	$	2.5    ^{	+0.1  }_{  -0.1 }	$	&  3 	&    927  $\pm$   64 	&  1034  $\pm$    41  &   8	&  $    0.4  ^{ 0.04  }_{   -0.04  }	$   	&  $      0.5  ^{   0.04 }_{   -0.04  }	$	&    89  $\pm$    3	&  14 \\
	NGC 4151       &    		 & 	$	6.0    ^{	+0.6  }_{  -0.2 }	$	&  3 	&   2680  $\pm$   64 	&  4711  $\pm$   750  &   9	&  $    8.4  ^{  0.9  }_{    -0.4  }	$   	&  $     26.0  ^{    6.4 }_{    -5.9  }	$	&    97  $\pm$    3	&  14 \\
	NGC 4253       &    		 & 	$	5.4    ^{	+0.2  }_{  -0.8 }	$	&  5 	&    538  $\pm$   82 	&   986  $\pm$   251  &   6	&  $    0.3  ^{  0.1  }_{    -0.1  }	$   	&  $      1.0  ^{    0.4 }_{    -0.4  }	$	&    93  $\pm$   32	&  16 \\
	NGC 4593       &    		 & 	$	4.5    ^{	+0.7  }_{  -0.6 }	$	&  3 	&   1561  $\pm$   55 	&  4141  $\pm$   416  &   10	&  $    2.1  ^{  0.3  }_{    -0.3  }	$   	&  $     15.1  ^{    3.2 }_{    -2.9  }	$	&   135  $\pm$    6	&  14 \\
	NGC 4748       &    		 & 	$	8.6    ^{	+0.6  }_{  -0.4 }	$	&  5 	&    791  $\pm$   80 	&  1373  $\pm$    86  &   6	&  $    1.1  ^{  0.2  }_{    -0.2  }	$   	&  $      3.2  ^{    0.4 }_{    -0.3  }	$	&   105  $\pm$   13	&  16 \\
	NGC 5273       &      	 & 	$	1.4    ^{	+1.1  }_{  -0.1 }	$	& 11 	&   1544  $\pm$   98 	&  4615  $\pm$   330  &   11	&  $    0.7  ^{  0.5  }_{    -0.1  }	$   	&  $      6.0  ^{    4.6 }_{    -0.9  }	$	&    74  $\pm$    4	&  20 \\
	NGC 5548       &     	 & 	$	5.5    ^{	+0.6  }_{  -0.7 }	$	&  5 	&   3900  $\pm$  266 	& 12539  $\pm$  1927  &   6	&  $   16.3  ^{  2.4  }_{    -2.6  }	$   	&  $    168.7  ^{   41.0 }_{   -42.5  }	$	&   195  $\pm$   13	&  16 \\
	NGC 6814       &      	 & 	$	7.4    ^{	+0.1  }_{  -0.1 }	$	&  5 	&   1697  $\pm$  224 	&  2945  $\pm$   283  &   6	&  $    4.2  ^{  0.8  }_{    -0.8  }	$   	&  $     12.5  ^{    1.7 }_{    -1.7  }	$	&    95  $\pm$    3	&  16 \\
	NGC 7469       &      	 & 	$	11.7   ^{	+0.5  }_{  -0.7 }	$	&  3 	&   1456  $\pm$  207 	&  2169  $\pm$   459  &   4	&  $    4.8  ^{  1.0  }_{    -1.0  }	$   	&  $     10.7  ^{    3.2 }_{    -3.3  }	$	&   131  $\pm$    5	&  14 \\
	PG 1411+442    &      	 & 	$	53.5   ^{ 	+13.1 }_{  -5.3 }	$	&  3 	&   1607  $\pm$  169 	&  2398  $\pm$   353  &   4	&  $   27.0  ^{  7.7  }_{    -4.8  }	$   	&  $     60.0  ^{   19.3 }_{   -13.8  }	$	&   209  $\pm$   30	&   5	\\
	PG 1426+015    &     	 & 	$	161.6  ^{   +6.9  }_{ -11.1 }	$	&  3 	&   3442  $\pm$  308 	&  6323  $\pm$  1295  &   4	&  $  373.6  ^{ 49.9  }_{   -53.8  }	$   	&  $   1260.8  ^{  369.1 }_{  -375.3  }	$	&   217  $\pm$   15	&  21 \\
	PG 2130+099    &      	 & 	$	31.0   ^{  	+4.0  }_{  -4.0 }	$	& 12 	&   1825  $\pm$   65 	&  2097  $\pm$   102  &   1	&  $   20.1  ^{  2.8  }_{    -2.8  }	$   	&  $     26.6  ^{    3.9 }_{    -3.9  }	$	&   163  $\pm$   19	&   5	\\
	SBS 1116+583A  &      	 & 	$	2.4    ^{ 	+0.9  }_{  -0.9 }	$	&  5 	&   1550  $\pm$  310 	&  3202  $\pm$  1127  &   6	&  $    1.1  ^{  0.5  }_{    -0.5  }	$   	&  $      4.8  ^{    3.0 }_{    -3.0  }	$	&    92  $\pm$    4	&  16 	
	\enddata
	\label{tab:rm_table}
	\tablenotetext{}{Ref. --- (1) Grier et al. 2012; (2) Dietrich et al. 2012; (3) Zu et al. 2011; (4) Peterson et al. 2004; (5) Grier et al. 2013b; (6) Park et al. 2012; (7) Barth et al. 2011; (8) Denney et al. 2010; (9) Bentz et al. 2006; (10) Denney et al. 2006; (11) Bentz et al. 2014; (12) Grier et al. 2013a; (13) Nelson \&\ Whittle 1995; (14) Nelson et al. 2004; (15) Woo et al. 2013; (16) Woo et al. 2010; (17) Ferrarese et al. 2001; (18) Kormendy \& Ho 2013; (19) Onken et al. 2004; (20) Cappellari et al. 2013; (21) Watson et al. 2008 
	\\}
	\tablenotetext{}{\bf Notes.}
	\tablenotetext{a}{\fwhb\ is measured from the rms spectrum in Barth et al. 2011.}
	\tablenotetext{b}{Only for this entry, \fwhb\ and \sigmahb\ are measured from the mean spectrum (Denney et al. 2010). All line width measurements except for this entry are measured from rms spectra.}
	
\end{deluxetable*}


\end{document}